\documentclass[preprint2]{aastex}

\slugcomment{submitted to Astrophys. J.}

\shorttitle{Dust Migration in Gas Disks}
\shortauthors{Takeuchi \& Artymowicz}

\begin{document}


\title{Dust Migration and Morphology in Optically Thin Circumstellar Gas
Disks\footnote{submitted to Astrophys. J.}}

\author{Taku Takeuchi and Pawel Artymowicz}
\affil{Stockholm Observatory, Stockholm University, SE-133 36
Saltsj\"obaden, Sweden}
\email{taku@astro.su.se, pawel@astro.su.se}

\begin{abstract}

We analyze the dynamics of gas-dust coupling in the presence of stellar
radiation pressure in circumstellar disks, which are in a transitional stage
between the gas-dominated, optically  thick, primordial nebulae, and the
dust-dominated, optically thin Vega-type disks. Meteoroids and dust undergo
radial migration, either leaving the disk due to a strong radiation pressure, or
seeking a stable equilibrium orbit in corotation with gas.  In our models of
A-type stars surrounded by a total gas  mass from  a fraction to dozens of
Earth masses,  the outward migration speed of dust is comparable
with the gas sound speed. Equilibrium orbits are circular, with exception
of those significantly affected by radiation pressure, which can be strongly
elliptic with apocenters extending beyond the bulk of the gas disk.

The migration of dust gives rise to radial fractionation of dust and creates a
variety of possible observed disk morphologies, which we compute by
considering  the equilibrium between the dust production (at radii $\lesssim
10$AU) and the dust-dust collisions  removing particles from their equilibrium 
orbits. Sand-sized and larger grains are distributed throughout most of the
disk, with concentration near the gas pressure maximum in the inner disk.
Smaller grains (typically in the range of 10 to $200 \micron$) concentrate in
a  prominent ring structure in the outer region of the gas disk (presumably at
radius $\sim$100 AU), where gas density is rapidly declining with radius. The
width and density, as well as density contrast of the dust ring with respect to
the inner dust disk depend on the distribution of gas and the mechanical 
strength of the particles, but do not depend on the overall  dust production
rate.

Our results open the prospect for deducing the distribution of gas in
circumstellar disks by observing their dust.   We have qualitatively  compared
our models with two observed transitional disks around HR 4796A and HD 141569A.
(Gas component has been detected, but not yet mapped  in detail, in the second
object.) Dust migration can result in observation of a ring or a  bimodal
radial dust distribution, possibly very similar to the ones produced by
gap-opening planet(s) embedded in the disk,  or shepherding it from inside or
outside. We conclude that  a convincing planet detection via dust imaging
should include specific non-axisymmetric structure (spiral  waves, streamers, 
resonant arcs) following from the dynamical simulations of perturbed disks.

\end{abstract}



\section{Introduction}

Dust and gas disks surround stars of different ages, from
pre-main sequence to post-main sequence stars. Because their evolution is
intimately connected with the process of planet formation, studies of dust
amount, size spectrum, mineralogy, and spatial distribution in disks can
provide valuable information about the crucial stages of planet formation.

The amount of dust in disks generally decreases with time (Zuckerman \&
Becklin 1993; Natta, Grinin, \& Mannings 2000). As a consequence, the fraction
of detectable, optically thick disks  (at $<$3 AU  from the star), decreases
from nearly 100\% at the estimated  object age $\sim$0.3 Myr down to a few
percent at the age of $\sim$10 Myr (Hillenbrand \& Meyer 1999; Meyer \& 
Beckwith 2000). The amount of gas decreases as well (Zuckerman, Foreville, \&
Kastner 1995; Liseau \& Artymowicz 1998), especially abruptly at an age of
several to $\sim$10 Myr for solar-type stars,  when circumstellar disks lose
most of the primordial protostellar/protoplanetary material through processes
such as  viscous accretion and photoionization, as discussed by Hollenbach et
al.\ (2000). From the original value of close to 100, the gas-to-solids (if
not necessarily gas-to-dust) mass ratio in the T Tauri and Herbig Ae/Be stars
decreases during this period (i.e., typically at age 3--10 Myr) to much
smaller values, eventually much less than unity, thus leaving a predominantly
dusty remnant disk. 

Vega-type stars, observable owing to their large thermal
infrared (IR) excess over the extrapolated IR stellar photospheric radiation,
and sometimes resolved in scattered visible/near-IR starlight, are thought to
contain such gas-poor disks. The prime example of an object of this kind is
$\beta$ Pic, which is 20--100 Myr old, and has a large disk ($>$10$^3$ AU
radius) containing gas and dust, apparently in a ratio smaller than unity
(Artymowicz 1997; Liseau \& Artymowicz 1998). Based on dust destruction
timescale in Vega-type disks, the disks are thought (or sometimes defined)
to be continuously replenished by the collisions and evaporation of
planetesimals.  Most convincingly in $\beta $ Pic, but also in a few younger,
spectroscopically variable Herbig Ae stars, the existence of planetesimals
and planetary objects necessary to perturb planetesimal orbits, has been
inferred from spectroscopy (Lagrange, Backman, \& Artymowicz 2000; Grady et
al.\ 2000).

The disks around stars with ages between 3 and 10 Myr  are particularly
interesting, because of the expected presence of protoplanets, growing by
collisional accumulation of planetesimals and through gas accretion
(Hayashi, Nakazawa, \& Nakagawa 1985; Lissauer 1993). We shall call such
disks transitional, in reference 
to a transition from a relatively high-mass, optically thick primordial
nebulae to low-mass, optically thin dust disks. (Other names, related to the
central star, are used for the objects in which such disks reside: young
main sequence stars, or old pre-main sequence stars; cf.\ Lagrange et al.\
2000.)

Recent imaging of two apparently transitional disks, HR 4796A and HD
141569A, provided interesting insight into the spatial distribution of
dust.  The A0 V star HR 4796A with age $8 \pm 2$ Myr (Stauffer, Hartmann, \&
Barrado y Navascues 1995) has been imaged, first in mid-IR
and afterwards in near-IR (Jayawardhana 1998; Koerner et
al. 1998; Schneider et al. 1999; Telesco
et al. 2000). The second object, B9.5 Ve star HD
141569A, is 2--10 Myr old.
Near-IR images of its dust disk were obtained  by 
Weinberger et al. (1999), and Augereau et al.
(1999a), while the thermal mid-IR flux was mapped by
Fisher et al. (2000). The disk in the image of HR 4796A
extends out to about 100 AU, a distance 4 times smaller than in HD
141569A, and is significantly narrower than its counterpart (of order 20 AU,
vs.\ 200 AU). The two systems have, however, very similar dust covering
factor, or the ratio of total IR dust luminosity to the stellar luminosity,
$f_{\rm dust}= L_{\rm dust}/L_*$, equal to 0.7\% in HR 4796A, and 1\% in
HD 141569A, which are a factor of 3 to 4 times larger than in $\beta$ Pic.

Both disks show structure, which can be modeled as axisymmetric, broad
rings with inner holes (near side-far side asymmetry of surface
brightness by a factor of 
1.5 or less, has been interpreted as
being due to an oblique viewing angle combined with anisotropic scattering
by dust.)
The radii of the inner holes are $\sim 60$ AU for HR 4796A (Koerner et
al. 1998; Schneider et al. 1999) and $\sim 150-200$ AU
for HD 141569A (Augereau et al. 1999a; Weinberger et
al. 1999). 
The image of HD 141569A taken at $1.1 \micron$ by Weinberger et
al. (1999) has been analyzed to 
reveal a double ring, or ring-gap-ring morphology with the shallow gap (or
dip) in the surface brightness, centered on the projected distance $r=250$
AU. The two maxima are found at 200 and 350 AU. Noteworthy, the same
system observed at $1.6 \micron$ by Augereau et al. (1999a) shows only one
broad ring peaking in brightness at $r=325$ AU. It is not clear to what
extent the apparent differences in the inferred surface brightness are due
to the peculiarities of observational techniques, such as the different 
time lag between the registration of object and comparison star 
(point-spread function), and to what extent by intrinsic factors like,
e.g.,  the possible wavelength-dependent scattering efficiency of dust 
around HD 141569A. For example, if the inner peak of HD 141569A is
dominated by small grains,  then it would Rayleigh-scatter 4.5 times better
at $1.1 \micron$ than at $1.6 \micron$. 
Recently, HD 163296, a Herbig Ae star, was announced by Grady et al.
(1999) to have a disk-and-gap morphology similar to that of HD 141569A.

Disks around Vega-type stars, which are more evolved than the
transitional disks discussed above, also have inner holes or clearing
regions, mostly but not always completely devoid of dust (for review see
Lagrange et al. 2000). 
Mid-IR images of the $\beta$ Pic disk show the central
low-density hole (Lagage \& Pantin 1994; Pantin, Lagage, \&
Artymowicz 1997) and   submillimeter images of Fomalhaut
and $\epsilon$ Eridani show the ring-like structure of dust disks (Holland
et al. 1998; Greaves et al. 1998). It is expected that the
gravity of embedded large bodies, such as planets or brown dwarfs,
form such inner holes in the dust distributions (Roques et al.
1994; Liou \& Zook 1999).

The inner holes of the transitional disks around HR 4796A and HD 141569A
may also be produced by planetary bodies embedded in the
disks (or rather, in these holes).
Moreover, planetary influences creating the disk ``gap''  at 325 AU
from HD 141569A (and a similar distance from HD 163296) have been proposed
as an explanation of these striking  features. Weinberger et al. (1999)
estimate that a 1.3 Jupiter mass  perturber would open a gap with the
observed extent.  Grady et al. (1999) conclude that their observations
lend strong support for early formation of massive planets at distances as
large as $r>300$ AU from the stars. 

However, there are several potential difficulties with the proposal that 
giant planets form routinely at such large distances in disks, in only a few 
Myr. Notice that the circular orbit of a planet, required by the circularity
of a narrow ``gap", would imply a local formation history of the hypothetical
planet. Formation in the inner disk and ejection to a large distance by,
e.g., planet-planet perturbations would not provide a nearly circular orbit
required. Admittedly, radial migration of planets via interaction with the
disk  may also occur but an {\em outward} migration distance is always
moderate.   Standard theory of planet formation requires a high surface
density of solids for giant planet formation at large radii. For example, the
numerical simulations of planetesimal accretion at 70 AU by Kenyon et al.
(1999) showed that the Pluto-sized planets (but not the required multi-Earth
mass planets) may form in 10 Myr only if the mass of the disk is as large as 
$10-20$ times the minimum solar nebula mass ($M_{\rm g} \sim 0.2-0.4
M_{\sun}$), a  mass about 10 times larger than the typical masses of disks
around T Tauri stars (Osterloh \& Beckwith 1995).

The amount of gas on the outskirts of a typical T Tauri disk,  at the age of 
$\sim $1 Myr, is probably also insufficient for other modes of giant planet
formation, such as the disk fragmentation, although such a possibility in
principle exists.  One argument against it, however, is apparent in the HD
141569A system, where the surface brightness profile does not show any deep
depletion  of dust (an empty gap) such as those which quickly appear  around a
giant planet in hydrodynamical models (e.g., Lubow, Seibert, \& Artymowicz
1999).  On the contrary, the ``gap'' is  merely a 14\% dip in the surface
density of dust, as compared with  neighboring disk regions. Weinberger et al.
(1999) thus propose  a very recent timing for the formation of their proposed
Jovian perturber. Troublesome as it is for  any formation scenario, this
special timing requirement  seems particularly incompatible with the early disk
fragmentation scenarios.

Motivated by the apparent axisymmetry or near-axisymmetry of  the observed
ring and gap structures in disks, we devote this paper to the exploratory 
study of an alternative mechanism for the appearance of a variety  of
possible dust morphologies including outer dust rings and apparent disk gaps,
connected with the phenomenon of radial  migration of dust in gas disks. 
In the context of transitional disks, which are optically thin to
the stellar visible and near-UV radiation and have gas component with
moderate mass, we consider two important effects: radiation pressure from the
central star and gas drag. 
As we will show, the combination  of radiation and drag forces
results in an {\em outward } direction of  migration for the observable dust
and strongly affects the  observed disk morphology, while preserving
the usual inward  direction of migration of sand and pebbles. It is easy to
see why transitional disks support the most vigorous migration.
Primordial protoplanetary nebulae are both optically thick and
so dense (gas:dust mass ratio $\sim 10^2$) that   most dust grains ``freeze"
in gas, and are not able to migrate through the gas. Vega-type stars, on the
other hand, are dust dominated (Lagrange et al. 2000). They  generally
contain so little gas  (gas:dust $\ll 1$) that the motion of dust is weakly
affected in the grain's lifetime. In contrast, HR 4796A and HD 141569A may
have gas disks comparable with or somewhat more massive than the dust disks.
Zuckerman et al. (1995) have found CO line emission from the gas inside 130 AU
distance from HD 141569A. A double H$\alpha$ line from the  gas disk in the
immediate vicinity of the star (Dunkin, Barlow, \& Ryan 1997) also 
demonstrates the existence of an extended gas disk, required for sustained
accretion of gas. Whether this gas is a remnant of primordial  gaseous nebula
or a secondary gas released by solids (for instance, evaporating
planetesimals), is not known at present. In any case, transitional disks may
have substantial amounts of gas  (many Earth masses),  even if an unfavorable
viewing geometry of most systems precludes a spectroscopic detection of the
circumstellar absorption lines. Artymowicz (2000) argues that gas must be
present and gas drag must be operative in the transitional disks, or else the
radiation pressure-boosted destruction rate  of solids would be prohibitively
large. 

One feature of our model, which may be observable, is the radial
fractionation of dust according to grain size.  The most important feature is
that  at the outer edge of the gas disk, dust forms a ring without the help
of  any shepherding bodies. Therefore, the ring structure of the dust disk
does not necessarily indicate the existence of planets or brown dwarfs.
Klahr \& Lin (2000) also proposed a model for the ring formation due to
the dust migration.
Whether or not these models can quantitatively  reproduce all the
observational data is, of course, a matter of detailed modeling, taking into
account each system's scale, central star's properties,  indications of dust
mineralogy from spectroscopy etc. We defer such detailed modeling of
transitional disks to a later publication, but provide qualitative
discussions in \S7.  It should, however, be stressed that our model is
testable against the existing observations, and can suggest new
observations.  Excluding dust migration as a cause of the
observed disk structure might, in fact, be as important as finding that it
matches the observations of a given object,  since such an exclusion would
very strongly support the hypothesis of dynamical sculpting by planets.  

The plan of the paper is as follows. 
In \S2, we describe the equations of the dust grain motion.
In \S3, we derive approximate expressions of radial velocity of grains.
We show both inward and outward migrations of grains occur due to the
combined effect of radiation pressure and gas drag forces.
In \S4, we show that grains segregate in the gas disk according to their
sizes and that grains of medium size ($10-100 \micron$ in our models)
concentrate at the edge of the gas disk.
In \S5, the orbital evolution of grains is calculated numerically.
We discuss the excitation of eccentricities of grain orbits at the edge 
of the gas disk.
In \S6, we calculate the lifetime of grains by taking into account the
collisional destruction of grains.
We derive the density structure of the dust disk and discuss the
formation of an outer dust ring.
In \S7, we show a sample image in near- and mid-IR from our simple model.
Finally, \S8 discusses the relevance of the migration-induced 
dust structures to observed systems, and compare them with dust features
of other origin.  


\section{Equations of Motion of Dust Grains in a Gas Disk}

We consider the motion of dust grains in a gas disk.
The equation of motion of a dust grain with mass $m$ and the position
vector ${\mbox{\boldmath $r$}}$ is
\begin{equation}
m \frac{d^2}{dt^2} {\mbox{\boldmath $r$}} = - \frac{G M m}{r^2}
{\mbox{\boldmath $\hat{r}$}} + {\mbox{\boldmath $F$}}_{\rm rad} +
{\mbox{\boldmath $F$}}_{\rm PR} + {\mbox{\boldmath $F$}}_{\rm g} \ ,
\end{equation}
where $r$ is the distance of the grain from the central star,
${\mbox{\boldmath $\hat{r}$}}$ is the unit vector directed to the grain,
$G$ is the
gravitational constant, $M$ is the central star's mass, and
${\mbox{\boldmath $F$}}_{\rm rad}$, ${\mbox{\boldmath $F$}}_{\rm PR}$, and 
${\mbox{\boldmath $F$}}_{\rm g}$ are the radiation pressure,
Poynting-Robertson drag (PR drag), and gas drag force, respectively.

The radiation pressure force ${\mbox{\boldmath $F$}}_{\rm rad}$ is
expressed using the ratio of radiation pressure force to the
gravitational force $\beta$,
\begin{equation}
{\mbox{\boldmath $F$}}_{\rm rad} = \beta \frac{G M m }{r^2}
{\mbox{\boldmath $\hat{r}$}} \ .
\end{equation}
The ratio ${\beta}$ is written as (Burns, Lamy, \& Soter 1979)
\begin{equation}
\beta = \frac{3 L Q_{\rm PR}}{16 \pi G M c s \rho_{d}} \ ,
\label{eq:beta}
\end{equation}
where $L$ is the central star's luminosity, $c$ is the speed of light, $s$
is the grain's radius, $\rho_{d}$ is the material density of the grain,
and $Q_{\rm PR}$ is the radiation pressure coefficient averaged over the
stellar spectrum.
The values of $\beta$ are calculated using Mie theory (Artymowicz
1988) and shown in Figure 1.
The grains are assumed to consist of silicates.  We use the laboratory 
measurements of optical constants of olivine (MgFeSiO$_4$) and
magnesium-rich pyroxene (Mg$_{0.8}$Fe$_{0.2}$SiO$_3$) by Dorschner et
al. (1995)\footnote{Optical data are available from
http://www.astro.uni-jena.de/Group/Subgroups/Labor/Labor/labor.html}. 
The grains are modeled as porous, containing 50\% volume fraction of vacuum
(the volume fractions of the two silicates being 25\% each).
The density of composite grains is then $\rho_{\rm d}=1.25 \ {\rm g \
cm^{-3}}$.
For the central stars, we used
parameters of HR 4796A (solid line) and HD 141569A (dashed line), and the
renormalized spectrum of A0 V type star Vega.
The mass and luminosity of HR 4796A are $M = 2.5 M_{\odot}$
and $L = 21.0 L_{\odot}$, respectively (Jura et al. 
1993; Jura et al. 1998), while for 
HD 141569A these parameters are  $M = 2.3 M_{\odot}$
and $L = 22.4 L_{\odot}$  (van den Ancker et al. 
1998).

The PR drag force ${\mbox{\boldmath $F$}}_{\rm PR}$ is 
\begin{equation}
{\mbox{\boldmath $F$}}_{\rm PR} = - \beta \frac{G M m }{r^2}
\left( \frac{v_{r}}{c}{\mbox{\boldmath $\hat{r}$}} + \frac{{\mbox{\boldmath
$v$}}}{c} \right) \ .
\end{equation}
where ${\mbox{\boldmath $v$}}$ is the velocity vector of the grain, and
$v_{r}$ is the velocity component in the ${\mbox{\boldmath $\hat{r}$}}$
direction (Burns et al. 1979). 

If the velocities of the grain
(${\mbox{\boldmath $v$}}$) and the gas (${\mbox{\boldmath $v$}}_{g}$)
differ, the
grain experiences a gas drag force ${\mbox{\boldmath $F$}}_{\rm g}$.
The sizes of grains considered in this paper are much smaller
than the mean free path of the gas molecules.
Thus, if the velocity difference ${\mbox{\boldmath $\Delta v$}} =
{\mbox{\boldmath $v$}} - {\mbox{\boldmath $v$}}_{\rm g}$ is much smaller
than the mean thermal velocity of gas, the gas drag force
is given by the Epstein drag law 
\begin{equation}
{\mbox{\boldmath $F$}}_{\rm g} = - \pi \rho_{\rm g} s^2  v_{\rm T}
\Delta {\mbox{\boldmath $v$}} \ ,
\end{equation}
where $\rho_{\rm g}$ is the gas density and $v_{\rm T}$ is 4/3 times the
mean thermal velocity.
For the gas with mean molecular weight $\mu_{\rm g}$ and temperature $T$,
$v_{\rm T}$ is 
\begin{equation}
v_{\rm T} = \frac{4}{3} \left( \frac{8 k T}{\pi \mu_{\rm g} m_{\rm H}}
\right)^{1/2} \ ,
\end{equation}
where $k$ is Boltzmann's constant and $m_{\rm H}$ is the mass of the
hydrogen atom.
In the case of $\Delta v = |\Delta {\mbox{\boldmath $v$}}| \gg v_{\rm
T}$, the gas drag force is
\begin{equation}
{\mbox{\boldmath $F$}}_{\rm g} = - \pi \rho_{\rm g} s^2  \Delta v \Delta
{\mbox{\boldmath $v$}} \ .
\end{equation}
A standard way to connect these two regimes is (Kwok 1975)
\begin{equation}
{\mbox{\boldmath $F$}}_{\rm g} = - \pi \rho_{\rm g} s^2  ( v_{\rm T}^2 +
\Delta v^2 )^{1/2} \Delta {\mbox{\boldmath $v$}} \ .
\end{equation}

The stopping time by the gas drag force is defined as $t_{\rm s} = 
{m \Delta v}/{|{\mbox{\boldmath $F$}}_{\rm g}|}$.
It can be made non-dimensional by writing  $t_{\rm s}= T_{\rm s} 
\Omega_{\rm K}^{-1}$, where $\Omega_{\rm K} = (G M / r^3)^{1/2}$ is
the Keplerian angular velocity.
The non-dimensional stopping time can then  be expressed as
\begin{equation}
T_{\rm s} =  \frac{T_{\rm ss}}{\sqrt{1+ \Delta v^2/ v_{\rm T}^2}} \ ,
\label{eq:Ts}
\end{equation}
where the square root factor represents the supersonic correction.
The subsonic stopping time parameter equals
\begin{equation}
T_{\rm ss}=\frac{4 \rho_{\rm d} s v_{\rm K}}{3\rho_{\rm g} r v_{\rm T}}
\ ,
\label{eq:Tss}
\end{equation}
where $v_{\rm K} = (G M / r)^{1/2}$ is the Keplerian velocity.
When $T_{\rm ss}=1$, particles
move slowly through the gas and 
couple to it in one dynamical time scale. 
Such particles are said to be marginally coupled to gas via gas drag, 
as opposed to the $T_{\rm ss}\gg 1$ case, representing large grains,
decoupled from the gas disk. 
Notice that to within a factor close to unity 
$\rho_{\rm g} r v_{\rm T}/
v_{\rm K}$ equals the gas surface density $\Sigma_{\rm g}$ in a disk.
This allows us to express $T_{\rm ss}$ as  $T_{\rm ss}\approx
\Sigma_{\rm 1p}/\Sigma_{\rm g}$,
the inverse ratio of $\Sigma_{\rm g}$ and a
``surface density of
one solid particle'', $\Sigma_{\rm 1p} =  
4 \rho_{\rm d} s/3$ (mass/cross-sectional area for a spherical particle). 
For example, a hypothetical 
uniform surface density gas disk with radius $100$ AU and 
total mass $3\times 10^{-5} M_\odot$, or 10 Earth masses, 
would have $\Sigma_{\rm g}= 8.5 \times 10^{-3}$ g/cm$^2$, 
the same as the $\Sigma_{\rm 1p}$ of an
$s=51\,\mu$m particle with density $\rho_{\rm d}=1.25$ g cm$^{-3}$.
All the dust grains smaller than about 51 $\mu$m would then be well coupled
to gas in such a disk, while all sand grains would have stopping times 
longer than one orbital period.

The velocity of gas $v_{\rm g}$ can be written using the 
Keplerian velocity $v_{\rm K}$ as
\begin{equation}
v_{\rm g} = v_{\rm K} ( 1 - \eta )^{1/2} \ .
\label{eq:omegag}
\end{equation}
Quantity $\eta$ is the ratio of the pressure gradient force to the
gravitational force, equal to 
\begin{equation}
\eta = - \frac{1}{r \Omega_{\rm K}^2 \rho_{\rm g}} \frac{dP_{\rm g}}{dr} \ ,
\label{eq:eta1}
\end{equation}
where $r$ is the distance from the central star, and $P_{\rm g}$ is the gas
pressure. If gas pressure decreases with radius, which 
normally happens everywhere in the disk except at the 
inner disk edge, $\eta >0$ and the gas orbital motion is slightly 
sub-Keplerian, that is slower than $v_{\rm K}$.

In this paper, we consider the dust particles near the mid-plane of the 
gas disk, where we introduce cylindrical coordinates $(r, \theta)$,
and neglect the fact that the grains might have 
finite orbital inclinations. 
This is equivalent to considering the
vertical motions to be averaged out by vertical integration over
an ensemble of particles in a short column
above and below the point $(r, \theta)$. 
This approach is justified by the generally 
weak coupling of vertical and planar motions of a grain in a thin disk.
The equations of motion of a dust grain at ${\mbox{\boldmath $r$}} =
(r, \theta)$ with velocity ${\mbox{\boldmath $v$}} = (v_r, v_{\theta})$
are
\begin{equation}
\frac{d v_r}{dt} = \frac{v_{\theta}^2}{r} - \frac{v_{\rm K}^2}{r} \left
[ 1 - \beta \left( 1 - \frac{2 v_r}{c} \right) 
\right] - \frac{v_{\rm K} v_r}{T_{\rm s} r} \ ,
\label{eq:motion_r}
\end{equation}
\begin{equation}
\frac{d}{dt} \left( r v_{\theta} \right) = 
 - \frac{\beta v_{\rm K}^2 v_{\theta}}{c} - \frac{v_{\rm K}}{T_{\rm s}} 
\left( v_{\theta} - v_{\rm g} \right) \ ,
\label{eq:motion_th}
\end{equation}
where $T_{\rm s}$ given by equation (\ref{eq:Ts}) depends on the
velocity of the grain.


\section{Radial Migration of Solids}

Small solid bodies migrate in disks with a speed depending on their size, 
radiative support and gas drag.  If the radiation pressure exceeds central
star's gravity ($\beta > 1$), then the grains (called $\beta$-meteoroids) are
blown away on a dynamical time scale, unless strongly braked by friction
against gas. We do not concern ourselves with the observability of  rapidly
escaping $\beta$-meteoroids in this paper. (We do take into  account their
collisions with stable disk particles, however.) In this section, we consider
grains that have nearly circular orbits and spiral in (or out) because of the
drag forces.

If the angular velocities of grains are larger than that of gas, the grains
experience headwind and lose angular momentum to move inward. Inward
migration of a grain subject to gas drag, but no radiation pressure or the PR
drag, was discussed by Adachi, Hayashi, \& Nakazawa (1976) and Weidenschilling
(1977). Exposure of dust grains to the radiation pressure often causes the
orbital speed of grains to be slower than that of gas. In this case which we
study below, the back-wind gives the grains angular momentum so that grains
move outward. Below, we study several regimes of migration allowing
analytical solutions of the equations of motion.

\subsection{Strongly Coupled Case ($T_{\rm s} \ll 1$)}

Consider the limiting case $T_{\rm s} \ll 1$,
in which the stopping time $t_{\rm s}$ is much smaller than the
dynamical time scale $\Omega_{\rm K}^{-1}$.
As will be clear from the results,  the particle moves azimuthally at nearly
the gas velocity, so that  $v_{\theta} \approx v_{\rm g}$. The grain
assumes a finite terminal radial velocity $v_{r}$  when the radial forces come
into balance so that $d v_r / dt = 0$. The terminal velocity  follows from
equation (\ref{eq:motion_r}),
\begin{equation}
v_{\rm g}^2 - v_{\rm K}^2 \left[ 1 - \beta \left(1 -\frac{2 v_{r}}{c}
\right) \right] - \frac{v_{\rm K} v_{r}}{T_{\rm s}} = 0 \ .
\label{eq:strong1}
\end{equation}
The term $2 v_{r} / c$ is negligible.
From equations (\ref{eq:Ts}) and (\ref{eq:omegag}), the above equation
becomes
\begin{equation}
v_{\rm K} ( \beta - \eta ) - \frac{1}{T_{\rm ss}} \left( 1 +
\frac{v_{r}^2}{v_{\rm T}^2} \right)^{1/2} v_{r} = 0 \ .
\label{eq:strong2}
\end{equation}
Solving this equation, we obtain
\begin{eqnarray}
v_{r} & = & {\rm sgn} ( \beta - \eta ) \frac{v_{\rm T}}{\sqrt{2}} \\
 & & \times \left[ \left\{ 1 + \left[2 T_{\rm ss} (\beta - \eta) \frac{v_{\rm
K}}{v_{\rm T}} \right]^2 \right\}^{1/2} - 1 \right]^{1/2} \ . \nonumber
\label{eq:vr_strong}
\end{eqnarray}
The particle migrates inward if $\beta < \eta$, or outward if $\beta >
\eta$.

The velocity of the grain in the azimuthal direction $v_{\theta}$ is
almost equal to the gas velocity $v_{\rm g}$, but there is a slight
difference between these velocity, $\Delta v_{\theta} = v_{\theta} -
v_{\rm g}$.
The angular momentum of the grain is approximated by
$H = m r v_{\rm g}$.  
The radial motion changes  the angular momentum secularly at a rate
\begin{equation}
\frac{dH}{dt} = \frac{m v_{\rm g} v_{r}}{2} \left( 1 - \frac{r}{1-\eta}
\frac{d \eta}{dr} \right) \ .
\end{equation}
The torque due to the gas and PR drag forces equals
\begin{equation}
\frac{dH}{dt} = - \frac{m}{T_{\rm ss}} \left( 1 + \frac{v_{r}^2 + \Delta
v_{\theta}^2}{v_{\rm T}^2} \right)^{1/2} v_{\rm K} \Delta v_{\theta} - m
\beta_{\rm c} v_{\rm K} v_{\rm g} \ ,
\end{equation}
where $\beta_{\rm c} = \beta v_{\rm K} / c$.
As seen from equation (\ref{eq:vt_strong}) below, $\Delta v_{\theta} \ll
v_{r}$.
Thus, $\Delta v_{\theta}^2$ in the parenthesis in the above equation can
be neglected. 
Equating the above two expressions for angular momentum change and
solving for $\Delta v_{\theta}$, the azimuthal wind velocity becomes
\begin{eqnarray}
\Delta v_{\theta} &=& - T_{\rm ss} \left( 1 +  \frac{v_{r}^2}{v_{\rm T}^2}
\right)^{-1/2} \\
 & & \times \left[ \frac{v_r}{2 v_{K}} \left( 1 - \frac{r}{1-\eta}
\frac{d \eta}{dr} 
\right) + \beta_{\rm c} \right] v_{\rm g} \ , \nonumber
\label{eq:vt_strong}
\end{eqnarray}
where $v_{r}$ is given by equation (\ref{eq:vr_strong}).
Notice that the formulae in this subsection remain valid in the 
supersonic migration regime, such as might arise from 
tight coupling of dust grain to gas, but even stronger coupling with stellar
radiation field. 

\subsection{Weakly Coupled Case ($T_{\rm s} \gg 1$)}

If $\beta > 1$, grains are blown away quickly by strong radiation
pressure force.
In the case of $\beta \gg 1$, the radial velocity of a grain was derived by
Lecavelier des Etangs, Vidal-Madjar, \& Ferlet (1998).
In this subsection, we consider the motion of a grain with $\beta < 1$.
The grain rotates with the speed $v_{\theta} = v_{\rm K} (1 -
\beta)^{1/2}$ and angular momentum $H = m r v_{\rm K} (1 -
\beta)^{1/2}$.
The velocity difference from the gas is $\Delta v_{\theta} = v_{\rm K} [(1 -
\beta)^{1/2} - (1 - \eta)^{1/2}]$.
The torque by the gas drag and PR drag forces is
\begin{equation}
\frac{dH}{dt} =  - \frac{m}{T_{\rm ss}} \left( 1 + \frac{v_{r}^2 + \Delta
v_{\theta}^2}{v_{\rm T}^2} \right)^{1/2} v_{\rm K} \Delta v_{\theta} - m
\beta_{\rm c} v_{\rm K} v_{\theta} \ .
\end{equation}
Because $v_{r} \ll \Delta v_{\theta}$ (see eq. [\ref{eq:vr_weak}]
below), $v_{r}^2$ in the right hand side is neglected.
The torque causes radial migration with velocity
\begin{equation}
v_{r} = \frac{2}{T_{\rm ss}} \left( 1 + \frac{\Delta
v_{\theta}^2}{v_{\rm T}^2} \right)^{1/2} \left[ \left( \frac{1
- \eta}{1 - \beta} \right)^{1/2} - 1 \right]
v_{\rm K} - 2 \beta_{\rm c} v_{\rm K} \ .
\label{eq:vr_weak}
\end{equation}
If PR drag force is negligible (the second term in
eq. [\ref{eq:vr_weak}] is neglected), a grain of $\beta < \eta$ migrates
inward, and a grain of $\beta > \eta$ migrates outward.

\subsection{Connecting Formulae For Arbitrary Stopping Time}

If the motions of grains and gas are close to the Keplerian circular
motion, we can derive approximate formulae connecting the two regimes
of weak and strong coupling of grains with gas.
Solutions of this kind,
clarifying the size-dependent nature of migration, 
were presented for the classical inward migration of grains
in optically thick disks by Adachi et al. (1976) and Weidenschilling
(1977).

We assume that $\eta$, $\Delta v_{\theta} / v_{\rm K}$, $v_{r} / v_{\rm
K}$, and $\beta_{\rm c} = \beta v_{\rm K} / c$ are much smaller than unity.
Based on the assumption that 
grains undergo a steady-state migration in a disk
without executing large radial 
motions in one turn of the orbit, we replace the full time derivatives
$d/dt$ in the equations of motion (\ref{eq:motion_r}) and
(\ref{eq:motion_th}) with the advective terms 
$v_r \partial /\partial r$. In other words, we set $\partial / \partial
t = 0$.
Then, neglecting the second order terms of $\eta$, $\Delta v_{\theta} /
v_{\rm K}$, $v_{r} / v_{\rm K}$, and $\beta_{\rm c}$,
the azimuthal equation (\ref{eq:motion_th}) becomes
\begin{equation}
\Delta v_{\theta} = - T_{\rm s} \left( \frac{v_{r}}{2} + \beta_{\rm c}
v_{\rm K} \right) \ .
\label{eq:vt_connect}
\end{equation}
The radial equation (\ref{eq:motion_r}) becomes
\begin{equation}
v_{\rm K} ( \beta - \eta) + 2 \Delta v_{\theta}
- \frac{1}{T_{\rm s}} v_{r} = 0 \ .
\label{eq:vr_connect0}
\end{equation}
Substituting equation (\ref{eq:vt_connect}) into equation
(\ref{eq:vr_connect0}), we obtain
\begin{equation}
v_{r} = \frac{\beta - \eta - 2 \beta_{\rm c} T_{\rm s}}{T_{\rm s} +
T_{\rm s}^{-1}} v_{\rm K} \ .
\label{eq:vr_connect}
\end{equation}
Note that the stopping time $T_{\rm s}$ defined by equation (\ref{eq:Ts})
is a function of $v_{r}$ and $\Delta v_{\theta}$.
Thus, $v_{r}$ and $\Delta v_{\theta}$ cannot be expressed analytically.
If the velocity difference of a grain with respect to 
gas is subsonic, $T_{\rm
s} \approx T_{\rm ss}$ is independent of the velocity of the grain.
In this case, $v_{r}$ and $\Delta v_{\theta}$ can be calculated
analytically by replacing $T_{\rm s}$ by $T_{\rm ss}$ in equations
(\ref{eq:vt_connect}) and (\ref{eq:vr_connect}).
In the supersonic regime, $v_{r}$ and $\Delta v_{\theta}$ must be
found numerically by an iterative correction method.

\subsection{Steady Orbits of Grains}

At some distance from the central star, the grain may have a steady
circular orbit without radial migration or any torque.
On such an orbit, $d v_r/dt = v_r = 0$ and $d v_{\theta} / dt = 0$.
From the $r$-component of equations of motion (\ref{eq:motion_r}), the
azimuthal velocity of the grain is 
\begin{equation}
v_{\theta} = v_{\rm K} (1 - \beta)^{1/2} \ .
\end{equation}
From the $\theta$-component of equations of motion (\ref{eq:motion_th}),
\begin{equation}
(v_{\theta} - v_{\rm g}) + \beta_{\rm c} T_{\rm s} v_{\theta} = 0 \ .
\label{eq:steady_orbit}
\end{equation}
The radius of the steady orbit is calculated as a solution of equation
(\ref{eq:steady_orbit}) for a given size of a grain $s$ (or $\beta$).
If PR drag force is negligible ($\beta_{\rm c} T_{\rm s} \ll 1$),
equation (\ref{eq:steady_orbit}) yields,
$v_{\theta} = v_{\rm g}$, and consequently
\begin{equation}
\beta = \eta \ .
\end{equation}
Thus, a grain has the steady circular orbit at such a distance
$r$ from the star, where $\beta = \eta(r)$.

\subsection{Size Dependence of Dust Particle Velocity}

In a gas disk, small grains of $\beta > \eta$ circulate faster than the  gas
and migrate outward, while large grains of $\beta < \eta$ rotate slower than
gas and migrate inward, if PR drag force is negligible.  Figure 2 shows the
dependence of the velocity of grains on their size. In this figure,
$\rho_{\rm g} = 2.18 \times 10^{-17}$ g cm$^{-3}$, $v_{\rm T} = 1.07 \times
10^5$ cm s$^{-1}$, and $\eta = 2.24 \times 10^{-2}$. These are the values the
disk in model 1, described below in \S4.1, has at 70 AU. The radial velocities
$v_{r}$ are shown by circles. The filled circles represent that grains move
outward ($v_{r} > 0$), while open circles represent that grains move inward
($v_{r} < 0$). These symbols are plotted by monitoring the velocities of
grains in the numerical calculations of their orbits, described
in \S5. The grains smaller than $174 \micron$ migrate outward, while larger
grains migrate inward. The velocity differences $\Delta v_{\theta}$ in the
azimuthal direction between grains and gas are shown by diamonds. The grains
smaller than $174 \micron$ rotate faster than gas, while larger grains rotate
slower. The $174 \micron$ grains have $\beta =\eta$, and thus 
steady circular orbits without radial migration. 

The grains with the size $0.1 \lesssim s \lesssim 10 \micron$ have similar
values of the radial velocity. These grains' $\beta$ are much larger than
$\eta$ and inversely proportional to the grain size $s$ (see Fig. 1). The
stopping time of these grains are much smaller than the orbital period
($T_{\rm s}=1$ at $s \approx 30 \micron$) and proportional to $s$ (eq.
[\ref{eq:Tss}]). It is seen that the radial velocity is constant in this
size range, following from equation (\ref{eq:vr_connect}). 

The solid lines show the approximate formulae (\ref{eq:vt_connect}) and
(\ref{eq:vr_connect}). The plotted values were obtained
numerically after several iterations for the supersonic correction. The
dashed lines show the first guesses of iteration, which are calculated using
the subsonic stopping time $T_{\rm ss}$ instead of $T_{\rm s}$.  The
deviations of the approximate and exact formulae
for the $0.1 \lesssim s \lesssim 10 \micron$ grains are due to the fact
that their radial velocities are comparable to the thermal speed of gas.


\section{Segregation of Dust According to Size}

As shown in \S 3, dust grains migrate inward or outward until they feel no
net torque.  The radius of the steady orbit is the function of grain's size
$s$, which is obtained as the solution of equation (\ref{eq:steady_orbit}).
Thus, in a gas disk, grains segregate according to their sizes. In this
section, we investigate the segregation of grains in a disk  with fixed gas
profile\footnote{ Gas profile will change significantly in time  if the total
gas mass is much smaller than the dust mass, a case which we do not consider
here. Gas migration and dust migration have  opposite directions and
magnitudes following from the conservation  of total angular momentum.}.

\subsection{Models of Gas Disks}

We assume that the temperature and density structures of the gas disk have
power law forms.
The temperature structure of the disk that is transparent to the central
star's visible radiation is (Hayashi 1981)
\begin{equation}
T = 278 \left( \frac{L}{L_{\sun}} \right)^{1/4} \left( \frac{r}{\rm 1
\ AU} \right)^{q} \ {\rm K} \ ,
\end{equation}
where power law index $q=-1/2$.
The mid-plane density has also power law, and at the outer radius of the
disk $r_{\rm out}$, the disk has an edge of characteristic
width $\Delta r_{\rm out}$.
The density structure is modeled by
\begin{equation}
\rho_{\rm g} = \rho_{0} \left( \frac{r}{\rm 1 \ AU} \right)^{p}
\frac{1}{2} \left( \tanh \frac{r_{\rm out} - r}{\Delta r_{\rm out}} + 1
\right) \ .
\end{equation}
In this model, the sound and thermal velocity $c_{\rm s}$ and $v_{\rm
T}$ are, respectively, 
\begin{eqnarray}
c_{\rm s} & = & \left( \frac{\Gamma k T}{\mu_{\rm g} m_{\rm H}}
\right)^{1/2} \nonumber \\
 & = & 1.17 \times 10^5 \left[ \frac{\Gamma}{1.4} \frac{2.34}{\mu_{\rm
g}} \left 
( \frac{L}{L_{\sun}} \right)^{1/4} \right]^{1/2} \nonumber \\
 & & \times \left( \frac{r}{\rm 1
\ AU} \right)^{q/2} {\rm cm \ s^{-1}} \ ,
\end{eqnarray}
\begin{eqnarray}
v_{\rm T} & = &\frac{4}{3} \left( \frac{8 k T}{\pi \mu_{\rm g} m_{\rm
H}} \right)^{1/2} \nonumber \\
 & = & 2.11 \times 10^5 \left[ \frac{2.34}{\mu_{\rm g}} \left
( \frac{L}{L_{\sun}} \right)^{1/4} \right]^{1/2} \nonumber \\
 & & \times \left( \frac{r}{\rm 1
\ AU} \right)^{q/2} {\rm cm \ s^{-1}} \ ,
\end{eqnarray}
where $\Gamma$ is the adiabatic exponent.
The ratio of the pressure gradient force to the gravitational force
$\eta$ is (eq. [\ref{eq:eta1}]) 
\begin{equation}
\eta = - \left( \frac{c_{\rm s}}{r \Omega_{\rm K}} \right)^2
\frac{1}{\Gamma} \left[ p + q - \frac{r \ {\rm sech}^2 x}{\Delta r_{\rm
out}  ( \tanh
x + 1)} \right] \ ,
\end{equation}
where $x = (r_{\rm out} - r) / \Delta r_{\rm out}$.

In model 1, we adopt following assumptions.
The disk gas consists of molecular hydrogen and atomic helium  
($\mu_{\rm g} = 2.34$ and $\Gamma = 1.4$).
The central star has the  mass and luminosity  of HR 4796A: 
$M = 2.5 M_{\sun}$ and $L = 21.0 L_{\sun}$, respectively.
The power law indexes of the density and temperature are $p = -2.25$ [which
corresponds to the power law index of surface density, $p_{\rm s} = p +
(q+3)/2 = -1.0$] and $q=-1/2$, respectively.
The outer radius of the gas disk is $r_{\rm out} = 100$ AU.
The width of the edge of gas disk is $\Delta r_{\rm out} = C_{\rm out}
(c_{\rm s} / r \Omega_{\rm K})_{r_{\rm out}} r_{\rm out}$.
The criterion for the Rayleigh stability at the edge of the disk
requires $C_{\rm out} \gtrsim 1$.
We adopt $C_{\rm out} = 1.05$, which yields $\Delta r_{\rm out} = 12.1 \
{\rm AU}$.
The mass of the gas disk is $M_{\rm g} = 10 M_{\earth}$ that corresponds
to $\rho_{0} = 3.12 \times 10^{-13} \ {\rm g \ cm^{-3}}$.
The dust grains have material density of $\rho_{\rm d} = 1.25 \ {\rm g \
cm^{-3}}$.
We also consider the other models of the gas disk:
smaller gas mass ($M_{\rm g} = 1 M_{\earth}$; model 2), steeper and
gentler density
gradient ($p = -2.75$, $p_{\rm s} = -1.5$; model 3, and $p = -1.25$,
$p_{\rm s} = 0$; model 4), and wider edge of the disk ($C_{\rm out} =
2.0$; model 5).
The values of model parameters are summarized in Table 1.

\subsection{Segregation of Grains}

Figure 3 shows the ratio of pressure gradient force to the gravitational
force $\eta$ as a function of radius $r$. The solid line shows $\eta$ for
model 1. For comparison, $\eta$ of the gas disk with a wider edge (model 5)
is shown as the dashed line. The value of the gas mass does not affect $\eta$
and model 2 ($M_{\rm g} = 1 M_{\earth}$) has the same profile of $\eta$ as
that in model 1. The profiles of $\eta$ in model 3 and 4 (different density
gradient; $p = -2.75$ and $-1.25$) are similar to that in model 1. In model
1 (solid line), $\eta$ increases gradually with radius to 80 AU, and
then more rapidly at the edge of the disk around 100 AU,
because of the steep decrease in the gas density and  pressure there.
A wider disk edge (cf,\ the dashed line) causes a more gradual rise of 
$\eta$.

The radius of the steady orbits of grains are calculated from equation
(\ref{eq:steady_orbit}) and are plotted in Figure 4 (we plot the grain size
$s$ as a function of radius $r$). In our models, the PR drag force is
negligible compared to the gas drag force except when the grains are at the
outside of the disk, $r \gtrsim 130 \ {\rm AU}$. The dotted line shows the
case in which the PR drag force is not included in model 1 and coincides with
solid line for $r \lesssim 130 \ {\rm AU}$. Thus, the PR drag force can be
neglected and the grains segregate to satisfy $\beta = \eta$. The smaller
grains stay at farther from the central star. At the edge of the disk where
$\eta$ varies rapidly ($80 \lesssim r \lesssim 110 \ {\rm AU}$), grains
in the size range $10 \lesssim s \lesssim 100 \micron$ find stable orbits.
Again, a wider edge in model 5 (dashed line) gives rise to a milder
concentration of grains.


\section{Orbital Evolution of Dust Grains}

In \S 3, we derived the radial velocities of grains analytically assuming the
orbits of the grains are nearly circular. In this section, we calculate the
orbital evolution of the grains numerically by integrating the equations of
motion (\ref{eq:motion_r}) and (\ref{eq:motion_th}). It will be shown that
the grains migrate to settle down in their own steady circular orbits, except
at the outer edge of the gas disk, where the eccentricities of the grains'
orbits are pumped up and dust disk extends beyond the gas
disk.

\subsection{Evolution of Orbital Radii and Eccentricities of Grains}

In Vega-type and transitional disks, dust grains are produced as ejecta 
of collisions between much larger 
parent bodies such as planetesimals, comets, or meteorites. These parent
bodies are not affected by radiation pressure and have Keplerian orbits. If
the velocity of ejection is much smaller than the orbital speed, the grains
produced by collisions have the same Keplerian velocities as their parent
bodies have. Because grains experience radiation pressure, the velocities
of the grains are larger than that the grain on the circular orbit would
have. Thus, the grains are endowed with orbital  eccentricities  
$e\approx  \beta / (1-\beta)$ (the equality becoming exact, 
if parent orbits are circular). 
However, if the density of gas is high enough that
the stopping time is much smaller than the orbital period, the
eccentricities of grains decrease quickly by the gas drag force and grains'
orbits are circularized. In the inner part of our model disks ($r \lesssim
20 \ {\rm AU}$), the orbits of grains with size $s \lesssim 100 \micron$
are circularized rapidly ($T_{\rm ss} < 1$).
Then, the grains spiral in or out as discussed in \S 3.

Figure 5 shows the evolution of the distance of grains from the central
star, $r$. The grains are assumed to be produced at 10 AU with eccentricity
$e = \beta / (1-\beta)$. These initial values are not important because the
grains rapidly forget their initial states due to the gas drag force. The
orbits of the grains are calculated by integrating the equations of motion
(\ref{eq:motion_r}) and (\ref{eq:motion_th}) numerically by 7th-order
Runge-Kutta method. The
integration are performed until $t=1000 P_{\rm out}$, where $P_{\rm out} =
2 \pi / \Omega_{\rm K} (r_{\rm out}) = 632$yr is the orbital period at $r_{\rm
out}(=100 \ {\rm AU})$.
Figure 5$a$ shows the evolution of orbits in model 1. The orbits of
the grains are quickly circularized and the grains migrate outward. The
grains larger than $100 \micron$ arrive at their steady orbits in $\sim 100
P_{\rm out}$, and settle down in these orbits. 
Smaller grains migrate to farther locations.
A grain of $20 \micron$ arrives at the radius of the steady orbit that
is in the outer edge of the 
disk in  $1 P_{\rm out}$, then its orbital eccentricity is pumped up. The
excitation of the eccentricity is seen in Figure 5$a$ as the oscillation of
the distance $r$. Finally, the orbital eccentricity of the $20 \micron$
grain decreases due to the gas drag force, and the grain stays in the
circular steady orbit. The orbital eccentricity of a grain of $8 \micron$
is also pumped up when the grain arrives at the edge of the disk, and then
decreases. Because the $8 \micron$ grain spends most of its time staying
outside the disk (it is in the gas disk only when being just around the
pericenter of the orbit), however, the gas drag force is not strong enough
to damp the eccentricity quickly. Thus, the $8 \micron$ grain stays in
eccentric orbit for long time (more than $1000 P_{\rm out}$). The evolution
of orbits in the disk with smaller gas mass ($M_{\rm g} = 1 M_{\earth}$;
model 2) is shown in Figure 5$b$. The main feature of the evolution is
similar to that in model 1. The settling time for grains to arrive at the
steady orbits is longer than that in model 1; $\sim 100P_{\rm out}$ for 100
and $300 \micron$ grains and $\sim 1000P_{\rm out}$ for $200 \micron$
grains. 

\subsection{Excitation of Eccentricity at the Disk's Outer Edge}

The orbital eccentricities of grains smaller than $20 \micron$ are excited
when these grains arrive at the outer edge of the gas disk. These grains have
large $\beta$ (0.19 for $20 \micron$ and 0.47 for $8 \micron$ grains). Until
the grains arrive at the disk edge, the gas drag force on such the small
grains are strong enough that the grains are trapped by the gas and move with
the gas. The gas rotates faster than the grains ($\eta=0.10$ at $r_{\rm
out}$), and the grains gain angular momentum from the gas and move outward.
Because the gas density decreases rapidly at the disk edge, the gas drag
force ceases suddenly when the grains pass through the disk edge. Therefore,
the grains are launched into the Kepler orbit (modified by the radiation
pressure) with initial velocity of ${\mbox{\boldmath $v$}}_{\rm g} = (0,
v_{\rm K}[1 - \eta]^{1/2})$ at the disk edge. With this initial velocity, the
orbits of the grains are eccentric with $e = (\beta - \eta)/ (1 - \beta)$ and
pericenters at the disk edge. After launching into the eccentric orbits, the
grains experience the gas drag force that works to circularize the grains'
orbits only when they are around their pericenter at the disk edge. Thus, the
eccentricities of the grains decreases slowly.

In the above discussion, we assumed that the gas density in the disk is
high enough for the gas drag force to hold the grains tightly with the gas.
We arrive at the same conclusion when the gas density is low and the gas drag
force is not strong enough to hold the grains.
Consider a small grain whose $\beta$ is much larger than $\eta$ at the
disk edge, and assume that the grain has slightly eccentric orbit.
The semi-major axis and eccentricity of its orbit are $a$ and $e$,
respectively.
Because $\beta \gg \eta$, the grain rotates much slower than the gas,
and experiences back-wind throughout its orbit, even when it is at the
pericenter.
The evolution of the eccentricity of the orbit by the perturbating drag
force is given by the Gauss's equation (Brouwer \& Clemence 1961)
\begin{eqnarray}
\frac{de}{dt} & = & \frac{1}{v_{\rm K}(a)}
\left( \frac{1-e^2}{1-\beta} \right)^{1/2} \\
 & & \times \left[ F_r \sin \psi + F_{\psi}
\left( \frac{\cos \psi + e}{1 + e \cos \psi} + \cos \psi
 \right) \right] \ , \nonumber
\label{eq:gauss_e}
\end{eqnarray}
where $\psi$ is true anomaly, and $F_r$ and $F_{\psi}$ are the radial
and azimuthal components of gas drag force, respectively.
At the pericenter ($\psi = 0$) and the apocenter ($\psi = \pi$),
equation (\ref{eq:gauss_e}) becomes
\begin{equation}
\frac{de}{dt} = \pm \frac{2}{v_{\rm K}(a)}
\left( \frac{1-e^2}{1-\beta} \right)^{1/2} F_{\psi} \ ,
\label{eq:gauss_e2}
\end{equation}
where plus and minus signs correspond to the pericenter and apocenter,
respectively. The grain experiences back-wind throughout the orbit. Thus,
$F_{\psi}$ is positive both at the pericenter and apocenter. From the
equation (\ref{eq:gauss_e2}), it is seen that the eccentricity increases at
the pericenter and decreases at the apocenter. When the grain is in the
edge of the disk, the gas density at the pericenter is much higher than
that at the apocenter. Therefore, the grain experiences stronger back-wind
at the pericenter and its eccentricity increases. The increase in the
eccentricity accompanies the increase in the velocity at the pericenter and
continues until the velocity arrives at the same velocity of the gas. When
the velocity of the grain at the pericenter becomes the same as that of the
gas, the eccentricity has grown up to be $e = (\beta - \eta)/ (1 - \beta)$.
After the eccentricity arrives at the maximum value, it starts to decrease.
This is because the gas drag force at the pericenter pushes the position of
the pericenter slightly outward. The eccentricity evolves to keep the
velocity of the grain at the pericenter being equal to the velocity of the
gas and maintains the value of $e = (\beta - \eta)/ (1 - \beta)$. Because
$\eta$ is an increasing function of $r$, the value of $\eta$ at the
pericenter increases as the pericenter moves outward, and the eccentricity
decreases. The gas density at the pericenter and the gas drag force
decrease as the pericenter moves outward, and the speed of the eccentricity
damping becomes more slowly.

Figure 6$a$ illustrates the evolution of the eccentricity and the
semi-major axis of an $8 \micron$ grain.
The average of the change in the eccentricity over one orbital period
$P$ is calculated by
\begin{equation}
\left\langle \frac{de}{dt} \right\rangle = \frac{1}{P} \int_{0}^{P}
\frac{de}{dt} dt \ .
\label{eq:gauss_e3}
\end{equation}
The integration is performed numerically and the effect of the PR drag
force is included.
The Gauss's equation of the variation in the semi-major axis is
\begin{eqnarray}
\frac{da}{dt} & = & \frac{2}{\Omega_{\rm K}(a) (1-\beta)^{1/2}
(1-e^2)^{1/2}} \nonumber \\
 & & \times \left[ F_r e \sin \psi + F_{\psi} \frac{a (1-e^2)}{r}
\right] \ .
\label{eq:gauss_a}
\end{eqnarray}

The average over one orbital period, $\langle da/dt \rangle$, is calculated
in analogy with equation (\ref{eq:gauss_e3}). The evolutions of $a$ and $e$
are shown in Figure 6$a$ as arrows. The length of the arrows are set to be
constant and does not represent the rate of the variation. A grain inside the
disk ($a \lesssim 100 {\rm AU}$) moves outward and its eccentricity
decreases. If the grain's pericenter is on the edge of the gas disk ($100
\lesssim a \lesssim 200 {\rm AU}$ and $0 < e \lesssim 0.6$), its eccentricity
grows. For a grain whose pericenter is outside the disk, the eccentricity and
semi-major axis decrease.  The evolution of the orbit of a grain whose
initial $a$ and $e$ are 100 AU and 0.05, respectively, is plotted as circles
on Figure 6$a$. The time interval of plotting circles is logarithmic: the
first time interval is $0.01 P_{\rm out}$ and each time interval is 1.2 times
the previous one. The last circle represents the orbit at $t=1000 P_{\rm
out}$. First, the grain's semi-major axis and eccentricity grow, then the
eccentricity turns to decrease. 

In Figure 6$b$, the evolution of orbital elements of a $100 \micron$
grain is
shown for comparison. It is seen that the eccentricity always decreases and
the semi-major axis changes toward the value of the steady orbit. Circles
show the evolutional path of a grain with the initial semi-major axis $a=100
{\rm AU}$ and eccentricity $e=0.5$. The eccentricity decreases quickly, and
the semi-major axis decrease at first, and then increases as the grain
approaches the steady orbit.

\subsection{Radial Distribution of Dust Grains}

The radial distribution of dust grains are obtained by the numerical
calculation described in \S5.1.
Figure 7 shows the positions of the pericenter and apocenter at $t=1000
P_{\rm out}$ against the grain size.
The solid line corresponds to model 1.
The grains smaller than $6.7 \micron$ are blown away (launched into
hyperbolic orbits when they move to outer edge of the gas disk).
The grains of $6.7 \le s \le 12 \micron$ have eccentric orbits ($e >
0.01$) with their pericenters on the outer edge of the gas disk.
The grains larger than $12 \micron$ have circular orbits.
The radii of the circular orbits
coincide to the radii calculated from the condition that the orbits are
steady (eq. [\ref{eq:steady_orbit}]) and shown as the dotted line.
The dashed line corresponds to model 2 (smaller gas mass; $M_{\rm g} =
1 M_{\earth}$).
Because the gas drag force is smaller than that in model 1, the
eccentricities of small grains remain larger.
The size of the smallest grain having circular orbit is $13 \micron$,
slightly larger than that in model 1.

In summary, the migration of dust grains produced at the inner part of the
gas disk proceeds as follows.
First, the grains migrate outward.
The large grains of $s > 12 \micron$ find the steady circular orbits in
the gas disk and settle down.
When the small grains ($s < 12 \micron$) arrive at the edge of the gas
disk, their eccentricities are excited.
The grains of $6.7 \le s \le 12 \micron$ have eccentric orbits, and make
dust disk extended outside the gas disk.
The grains small enough for their eccentricities to be pumped up to $e >
1$ are blown away.
The grains of $10 \lesssim s \lesssim 100 \micron$ are concentrated at
the edge of the gas disk.
It is expected that these grains make a dust ring at the edge of the gas
disk.
In the next section, we estimate the number density of grains in the
ring.


\section{Lifetime and Density of Dust Grains}

The grains larger than about $10 \micron$ have steady circular
orbits in the gas disk and segregate according to their sizes.
The size of grains in the steady orbit is function of distance from the
star $r$, written as
\begin{equation}
s_{\rm s} = s_{\rm s} (r) \ .
\end{equation}
The grains of size $s_{\rm s}$ in the steady orbit do not collide each
other, because their orbits are circular.
However, smaller grains ($s < s_{\rm s}$) whose steady orbits are farther
from the star flow outward and cross the orbit of grains of size $s_{\rm
s}$.
These smaller grains collide and destroy the grains in the steady orbit.
In this section, we calculate the lifetime of grains in the steady
circular orbit in the gas disk, then estimate their number density.
The small grains ($s \lesssim 10 \micron$) extended outside the gas disk
have eccentric orbits and collide each other.
The calculation of the lifetime of these small grains remains as future
work.

The grains are originally produced by collisions of parent bodies such
as planetesimals or comets.
In this section, we assume that these
parent bodies are distributed in the innermost part of the gas disk
(say, $\sim 10 {\rm AU}$).
The grains produced by collisions migrate outward to settle down in the
steady orbits.
When the grain on the steady orbit is broken up by collisions, the
fragments flow outward and settle down in the new steady orbits.
Thus, grains gradually migrate outward every collisional destructions.
The steady orbit of grains of size $10 \lesssim s_{\rm s} \lesssim 100
\micron$ are concentrated at the edge of the gas disk (Fig. 4).
Thus, it is expected that the grains of $10 \lesssim s_{\rm s} \lesssim 100
\micron$ gather in the edge of the gas disk and make a dust ring.
The grains are finally blown away when collisional destruction reduces
their sizes so small that they cannot have steady orbit in the gas disk.

\subsection{Collisional Destruction of Grains}

There are two phases of collisional destruction.
A grain is broken up by a collision with large impact kinetic energy
(catastrophic disruption), while part of a grain is ejected by a
collision with small energy (cratering).
Consider the destruction of a grain of mass $m = 4 \pi \rho_{\rm d} s^3
/ 3$ by the collision with a grain of mass $m'$.
The mass $\Delta m$ lost from the target grain $m$ by cratering with the
collisional velocity $u$ is (Holsapple 1993)
\begin{equation}
\frac{\Delta m}{m} = K \frac{m'}{m} \left( \frac{u}{\rm cm \ s^{-1}}
\right)^{3 \mu} \ ,
\label{eq:creter}
\end{equation}
where $K$ is a coefficient representing efficiency of cratering and the
value of the exponent $\mu$ is uncertain now.
We assume that the mass loss $\Delta m$ is proportional
to the impact energy and adopt $\mu = 2/3$.
In equation (\ref{eq:creter}), $\Delta m / m$ becomes larger than 1 for
large $u$, that corresponds to catastrophic disruption.
Of course, $\Delta m / m$ cannot be larger than 1 and should be
\begin{equation}
\frac{\Delta m}{m} = \max \left\{ K \frac{m'}{m} \left( \frac{u}{\rm cm \
s^{-1}} \right)^{2} \ , \ 1 \right\} \ .
\label{eq:massloss}
\end{equation}
The value of $K$ is calculated from the minimum energy for the catastrophic
disruption.
Let $Q^*$ be the threshold of the specific energy per unit target mass
for the catastrophic disruption.
The threshold energy $Q^*$ depends on the impact velocity $u$ and target
size $s$, and written as (Housen \& Holsapple 1999)
\begin{equation}
Q^* = Q^*_0 \left( \frac{u}{\rm km \ s^{-1}} \right)^{2-3\mu}
\left( \frac{s}{10 \ \rm cm} \right)^{\delta} = Q^*_0 \left( \frac{s}{10
\ \rm cm} \right)^{\delta}\ ,
\label{eq:Qstar}
\end{equation}
where we used $\mu = 2/3$.
The values of $Q^*_0$ are obtained by the experiments: $Q^*_0 = 10^7
\ {\rm erg \ g^{-1}}$ for rocks (Housen \& Holsapple 
1999) and $Q^*_0 = 4 \times 10^5 \ {\rm erg \ g^{-1}}$ for ice (Fig. 9 in
Arakawa 1999).
The value of the exponent $\delta$ is uncertain.
Housen \& Holsapple (1990) proposed that $\delta =
-0.24$ and Housen \& Holsapple (1999) recommended that 
$\delta = -0.67$.
We are interested in the lifetime and density of grains constituting the
dust ring at the edge of the gas disk.
The sizes of these grains are $10 \lesssim s \lesssim 100 \micron$.
Thus, we estimate $Q^*$ for the grain of size $s = 50 \micron$ as
\begin{equation}
Q^* = \cases{
6.2 \times 10^7 - 1.6 \times 10^9 \ {\rm erg \ g^{-1}} & for rocks \cr
2.5 \times 10^6 - 6.5 \times 10^7 \ {\rm erg \ g^{-1}} & for ice \cr
} \ ,
\end{equation}
where smaller values correspond to $\delta =
-0.24$ and larger values correspond to $\delta = -0.67$.
The definition of the catastrophic disruption is $\Delta m / m \ge 1/2$,
which yields
\begin{equation}
K = \frac{1}{4 Q^*} = \cases{
1.5 \times 10^{-10} - 4.0 \times 10^{-9} & for rocks \cr
3.8 \times 10^{-9} - 1.0 \times 10^{-7} & for ice \cr
} \ .
\label{eq:effdest}
\end{equation}

\subsection{Lifetime of Grains}

In this subsection, we derive the equation for the life time of a grain
of size $s_{\rm s}$ on the steady orbit.
A grain $s_{\rm s}$ collide with smaller grains crossing its orbit.
The probability of collision with grains in the size range $[s',
s'+ds']$ in $1 \ {\rm sec}$ is
\begin{equation}
p_{\rm col} = \sigma u n' ds' \ ,
\end{equation}
where $\sigma = \pi (s_{\rm s} + s')^2$ is the collisional cross
section, $n' ds'$ is the number density of grains in the size range $[s',
s'+ds']$, and $u = |{\mbox{\boldmath $v'$}} - {\mbox{\boldmath $v$}}|$
is the relative velocity of the grains.
The velocity of the grain $s_{\rm s}$ is the same as the velocity of the
gas, i.e., ${\mbox{\boldmath $v$}} = (0, v_{\rm g})$.
The velocity ${\mbox{\boldmath $v'$}} = (v'_{r}, v'_{\theta})$ of the
grain $s'$ is obtained by the orbital calculations of grains described in
\S5.1.
For small grains in the inner part of the disk where the gas density is
high, the non-dimensional stopping time $T_{\rm s}$ is much smaller than
unity.
We use the analytical formulae (\ref{eq:vr_strong}) and
(\ref{eq:vt_strong}) to calculate the velocities of small grains if
$T_{\rm s} \ll 1$, because numerical calculations of orbits consume much
time.
The density $n' ds'$ is calculated from the production rate of dust
grains.
Let the number of grains that are in the size range $[s', s'+ds']$ and cross
the circle of the radius $r$ (the orbit of the grain $s_{\rm s}$) in $1
\ {\rm sec}$ be $\dot{N}_{\rm pro}(r, s') ds'$.
The number density $\dot{N}_{\rm pro}(r, s')$ is assumed to have power law
form in $s'$. 
We also assume that the grains are originally produced innermost part of
the gas disk and flow outward. 
Only the grains smaller than $s_{\rm s}$ cross the orbit of the grain
$s_{\rm s}$.
The grains larger than $s_{\rm s}$ stay in their steady orbits which are
closer to the star, but when they are broken to pieces smaller than $s_{\rm
s}$, these pieces flow outward and cross the orbit of the grain $s_{\rm
s}$.
Thus, the grains crossing the orbit of the grain $s_{\rm s}$ are
constituted of smaller grains, and $\dot{N}_{\rm pro}(r, s') ds'$ are written as
\begin{equation}
\dot{N}_{\rm pro}(r, s') ds'= \cases{
\dot{N}_{0}(r) \left( \frac{s'}{{\rm cm}} \right)^{\alpha} ds' & for $s
\le s_{\rm s}$ \cr
0 & for $s > s_{\rm s}$ \cr
} \ .
\label{eq:numflux}
\end{equation}
The mass of grains crossing the orbit of the grain $s_{\rm s}$ in $1 \ 
{\rm sec}$ is
\begin{equation}
\dot{M}_{\rm pro} = \int_{s_{\rm min}}^{s_{\rm s}} \frac{4 \pi}{3} s^{\prime
3} \rho_{\rm d} \dot{N}_0 \left( \frac{s'}{{\rm cm}} \right)^{\alpha} ds' \ ,
\label{eq:masspro}
\end{equation}
where $s_{\rm min}$ is the minimum size of grains.
In the steady state, $\dot{M}_{\rm pro}$ is independent on $r$ and represents
the mass production rate of grains in the gas disk.
From equation (\ref{eq:masspro}),
\begin{equation}
\dot{N}_{0}(r) = \frac{3 \dot{M}_{\rm pro}}{4 \pi \rho_{\rm d}} \left[ \int_{s_{\rm
min}}^{s_{\rm s}} s^{\prime 3} \left( \frac{s'}{{\rm cm}}
\right)^{\alpha} ds' \right]^{-1} \ .
\end{equation}
Note that $\dot{N}_{0}$ is the function of $r$ because $s_{\rm s}$ is the
function of $r$.
The other expression of number flux is $\dot{N}_{\rm pro}(r, s') ds' = 2
\pi r h_{\rm d} v'_{r} n' ds'$, where $h_{\rm d}$ is the thickness of
the dust disk.
Solving this equation and equation (\ref{eq:numflux}) gives
\begin{equation}
n' ds'= \cases{
\frac{\dot{N}_{0}}{2 \pi r h_{\rm d} v'_{r}} \left( \frac{s'}{{\rm cm}}
\right)^{\alpha} ds' & for $s \le s_{\rm s}$ \cr
0 & for $s > s_{\rm s}$ \cr
} \ .
\label{eq:dustden}
\end{equation}
The mass loss rate from the grain $s_{\rm s}$ with mass $m$ is
\begin{eqnarray}
\frac{1}{m} \frac{dm}{dt} & = & \int_{s_{\rm min}}^{s_{\rm s}}
\frac{\Delta m}{m} \sigma u n' ds' \nonumber \\
& = & \frac{\dot{N}_0}{2} \left( \frac{h_{\rm d}}{r} \right)^{-1} \\
& & \times \int_{s_{\rm
min}}^{s_{\rm s}} \frac{\Delta m}{m} \frac{u}{v'_{r}} \left
( \frac{s_{\rm s}+s'}{r} \right)^2 \left( \frac{s'}{{\rm cm}} \right)^{\alpha}
ds' \ . \nonumber 
\end{eqnarray}
The life time of the grain $s_{\rm s}$ is
\begin{eqnarray}
t_{\rm life} & = &  m \left( \frac{dm}{dt} \right)^{-1} \nonumber \\
& = & 2 \dot{N}_0^{-1} \frac{h_{\rm d}}{r} \\
& & \times \left[ \int_{s_{\rm
min}}^{s_{\rm s}} \frac{\Delta m}{m} \frac{u}{v'_{r}} \left
( \frac{s_{\rm s}+s'}{r} \right)^2 \left( \frac{s'}{{\rm cm}} \right)^{\alpha}
ds' \right]^{-1} \ . \nonumber 
\end{eqnarray}

\subsection{Spatial Density of Grains}

Consider an annular part of the disk between $r$ and $r+dr$.
The annulus is made of the grains of the size $[s_{\rm s}, s_{\rm s} +
(d s_{\rm s} / dr) dr]$.
The total number of the grains in the annulus is
\begin{equation}
N_{\rm d} ds_{\rm s} = 2 \pi r h_{\rm d} n_{\rm s} dr \ ,
\end{equation}
where $n_{\rm s}$ is the number density of the grains in the steady
orbits\footnote{Note that the unit of $n_{\rm s}$ is cm$^{-3}$, while
that of $n'$ is cm$^{-4}$.}.
The number of grains destroyed by collisions in 1 sec is
\begin{equation}
\dot{N}_{\rm des} ds_{\rm s} = \frac{N_{\rm d} ds_{\rm s}}{t_{\rm life}} =
\frac{2 \pi r h_{\rm d} n_{\rm s} dr}{t_{\rm life}} \ .
\end{equation}
The supply of the grains to the annulus is $\dot{N}_{\rm pro}(r, s_{\rm s})
ds_{\rm s} = \dot{N}_{0} (s_{\rm s}/{\rm cm})^{\alpha} ds_{\rm s}$.
In the steady state, the rate of destruction and supply balance each
other.
From $\dot{N}_{\rm des} ds_{\rm s} = \dot{N}_{\rm pro}(r, s_{\rm s})
ds_{\rm s}$, we obtain
\begin{eqnarray}
n_{\rm s} & = & \frac{1}{\pi r^2} \left( \frac{s_{\rm s}}{{\rm cm}}
\right)^{\alpha} \frac{ds_{\rm s}}{dr} \\
& & \times \left[ \int_{s_{\rm
min}}^{s_{\rm s}} \frac{\Delta m}{m} \frac{u}{v'_{r}} \left
( \frac{s_{\rm s}+s'}{r} \right)^2 \left( \frac{s'}{{\rm cm}} \right)^{\alpha}
ds' \right]^{-1} \ . \nonumber 
\end{eqnarray}
We define the optical depth of the dust disk in the vertical direction
using the geometrical cross section as
\begin{eqnarray}
\tau & = & \pi s_{\rm s}^2 h_{\rm d} n_{\rm s} \nonumber \\
& = & \frac{s_{\rm s} h_{\rm d}}{r^2} \left( \frac{s_{\rm s}}{{\rm cm}}
\right)^{\alpha+1} \frac{ds_{\rm s}}{dr} \\
& & \times \left[ \int_{s_{\rm
min}}^{s_{\rm s}} \frac{\Delta m}{m} \frac{u}{v'_{r}} \left
( \frac{s_{\rm s}+s'}{r} \right)^2 \left( \frac{s'}{{\rm cm}} \right)^{\alpha}
\frac{ds'}{{\rm cm}} \right]^{-1} \ . \nonumber
\end{eqnarray}
Note that the density and optical depth of the grains are independent on
$\dot{N}_0$ or the production rate of grains $\dot{M}_{\rm pro}$.

\subsection{Results of Numerical Calculations}

In this subsection, we calculate the life time of grains and density
(optical depth) of the dust disk in model 1. We assume that the grains are
made of rocky material and adopt $K = 1 \times 10^{-9}$. The mass
production rate of grains are uncertain. The models of dust disks (however,
without gas disks) suggested that the rate of mass loss from the disk is
of order $10^{-6} M_{\earth} {\rm yr}^{-1}$ for $\beta$ Pic (Artymowicz \&
Clampin 1997), and HR 4796A (Augereau et al.
1999b). In the steady state, the mass loss rate balances
with the mass production rate. We adopt mass production rate $\dot{M}_{\rm pro} =
10^{-6} M_{\earth} {\rm yr}^{-1}$. The density of grains or the optical
depth of the disk do not depend on the mass production rate, while the life
time of grains is inversely proportional to it. The thickness of the dust
disk is assumed to be $h_{\rm d} = 0.1r$ and the minimum size of grains
$s_{\rm min} = 0.01 \micron$ is adopted. The power law index of $\dot{N}_{\rm
pro}$ in the equation (\ref{eq:numflux}) is assumed to be $\alpha=-3.5$,
a typical value usually assumed for collisional cascade.

The solid line in Figure 8$a$ shows the life time of grains in model 1 (gas
mass $M_{\rm g} = 10 M_{\earth}$). The grains larger than $100 \micron$ are
in the inner part of the gas disk. Their life time is order of 1000
times the orbital period $P_{\rm out}$ at $r_{\rm out}(= 100 {\rm AU})$ and
decreases with the size because larger grains inhabit inner part 
of the disk where the density of colliding grains is higher (eq.
[\ref{eq:dustden}]). The grains less than $100 \micron$ are in the edge of
the gas disk. At the edge of the disk, the collisional velocity of grains
becomes large because the gas density is not high enough for the gas drag
force to hold the grains in the same velocity. Thus, the life time of these
small grains is much shorter than that of larger grains ($s \gtrsim 100
\micron$) in the gas disk. In model 1, the life time of grains is much
larger than the migration time to the steady orbits (the dashed line in
Fig. 8$a$). Thus, the grains spend most of their time in their steady
orbit and the dust disk is mainly composed of the grains in the steady
orbit. The contribution of grains flowing toward the steady orbit to the
density of the dust disk or the optical depth can be neglected.

The solid line in Figure 9$a$ shows the optical depth of the dust disk in
model 1. The dominant profile of the optical depth is the peak at the edge of
the gas disk ($\sim 90 {\rm AU}$). This peak is made of the grains of $10
\lesssim s_{\rm s} \lesssim 100 \micron$. Because these grains gather in
the edge of the gas disk (see Fig. 4), the density of grains becomes high
and a dust ring forms there. The life time of grains outside the gas disk
($r \gtrsim 100 {\rm AU}$ and $s \lesssim 10 \micron$) is much shorter than
that of grains in the gas disk. Thus, the optical depth declines steeply
outside the gas disk. The large grains ($s_{\rm s} \lesssim 300 \micron$)
condense at the innermost part of the gas disk. The density of gas is high
at the inner part of the disk, and strong gas drag force suppresses the
collisional disruption of grains. Thus, the optical depth at small radius
becomes large and an inner dust disk forms at $r \lesssim 40$AU.

\subsection{Models of Dust Distribution}

The structure of dust disks strongly depends on the properties of gas disks.
In this subsection, we show how the structure of the dust disk varies
with the variation of the gas disk.

\subsubsection{Small Gas Mass}

The solid line in Figure 8$b$ shows the life time of the grains in the gas
disk with smaller mass (model 2; $M_{\rm g} = 1 M_{\earth}$). The life time
of the grains is much shorter than in model 1, because the weaker gas drag
force results in high collisional velocity of grains. In model 2, the life
time is comparable or smaller than the time for the grains to migrate to 
their steady orbits (the dashed line in Fig. 8$b$). This means that the
dust disk is
composed of both the grains in the steady orbits and the grains flowing
outward to the steady orbits. Some grains flowing outward are destroyed
before arriving at the steady orbit. In this case, the calculation of the
density of grains described in \S6.3 is not appropriate. The ring structure
at the edge of the gas disk would be less prominent, because some of grains
making the ring would be destroyed before arriving at their steady orbit.
To obtain the structure of the dust disk, the calculation taking into
account the destruction of grains on the way to the steady orbit must be
needed.  If production rate of grains is 10 times smaller and $\dot{M}_{\rm pro}
= 10^{-7} M_{\earth} {\rm yr}^{-1}$, the life time of grains is 10 times
longer, and our calculation of density would be appropriate. The optical
depth in the model of  $M_{\rm g} = 1 M_{\earth}$ is shown as the dashed
line in Figure 9$a$, assuming that the life time of grains are much larger
than the migration time. The profile of the optical depth is similar to that
in the model of $M_{\rm g} = 10 M_{\earth}$, while its magnitude is about
10 times smaller. At the peak of the optical depth in the dust ring ($r
\approx 90 {\rm AU}$), $\tau = 1.1 \times 10^{-1}$ for $M_{\rm g} = 10
M_{\earth}$ and 
$\tau = 1.2 \times 10^{-2}$ for $M_{\rm g} = 1 M_{\earth}$.

\subsubsection{Different Gas Profiles}

In Figure 9$b$, we show the optical depth of the dust disks for
various density gradients of gas disks. The dashed line shows the disk with
steeper density gradient (the power law index of the gas density $p=-2.75$
corresponding to that of the surface density $p_{\rm
s}=-1.5$; model 3). The mass of the gas disk is the same as in model 1
($M_{\rm 
g} = 10 M_{\earth}$). Because of the steeper density gradient, the gas
density in the inner part of the disk is higher than that in model 1, while
the gas density at the edge of the disk is lower. The lower gas density at
the edge of the disk causes the shorter life time of grains and the smaller
optical depth of the dust ring. The grains in the inner part of the disk
are strongly held by dense gas.
The collisional velocity of these grains are
small and collisional destruction of grains is not effective. The grains in
the inner part of the disk survive longer than the grains in model 1, and
the optical depth increases with decreasing radius $r$. The inner optically
thick disk may obscure the star from the outer part of the disk. In this
case, the grains in the outer part of the disk do not experience the
radiation pressure force and do not have steady orbits, so that the dust
ring would disappear.

The dotted line in Figure 9$b$ shows the disk with the constant surface
density ($p=-1.25$ and $p_{\rm s}=0$; model 4).  The gas density in the
inner part of the disk is much lower than that in the model 1, while the
gas density at the edge of the disk is higher. The optical depth of the
dust ring is larger than in the model 1 because of the higher gas density
at the edge of the disk. In the gas disk with the constant surface density,
even the grains in the innermost part of the disk are not strongly held by
gas and are destroyed in short time by collisions with high velocity. The
number density of these grains is not as large as that in model 1 or 3. 
The inner disk is optically thin.
In model 4, because the optical depth decreases with decreasing radius
$r$, the inner disk does not obscure the star.

In the previous models, we assumed that the gas disks have sharp outer
edges
(marginally Rayleigh stable at the edge where $C_{\rm out}=1.05$). The
sharpness of the edge is not known well. Recent images of the dust disk of
GG Tau show a very sharp edge ($\Delta r_{\rm out} / r_{\rm out} \sim
0.04$; Guilloteau, Dutrey, \& Simon 1999). However, the
edge of gas disk may be smooth. The optical depth of the dust disk in the
gas disk with smooth edge ($C_{\rm out}=2.0$; model 5) is shown in Figure
9$c$. The concentration of grains at the disk edge is moderate, and the
ring feature is weak. 

\subsubsection{Grains of Smaller Collisional Strength}

The value of the efficiency of collisional destruction $K$ is uncertain
(eq. [\ref{eq:effdest}]). If the grains are stronger (weaker), their
life time is longer (shorter) and their density is higher (lower). As an
example, we show the optical depth of the dust disk with $K=2\times
10^{-8}$, which is 20 times larger than $K$ in model 1 and corresponds to
the efficiency of collisional destruction of icy grains (model 6), in
Figure 9$d$. The life time of the grains is about 20 times shorter than
that in model 1, and the optical depth is also 20 times smaller. The icy
grains are destroyed more easily than the rocky grains and are blown away
by the radiation pressure more quickly. Therefore, the dust disk are mainly
composed of rocky grains with little icy mantle. This is consistent with
the properties of the grains derived through the spectral fitting of
thermal emission from the HR 4796A dust ring by Augereau et al.
(1999b).


\section{Morphology of the Ringed Systems}

We consider first the global quantities (mass and luminosity) of the 
dust, as well as gas, in the models, and compare them with some observed
quantities. Next, we discuss the morphology of the model disks
in the thermal and scattered light images. 

\subsection{Mass and Luminosity of Model Dust Rings}

The mass of the dust disk $M_{\rm d}$ is given by 
\begin{equation} 
M_{\rm d} = 
\frac{8}{3} \pi^2 \rho_{\rm d} \int_{r_{\rm in}}^{r_{\rm out}} r
h_{\rm d} s_{\rm s}^3 n_{\rm s} dr \ . 
\end{equation} 
We set $r_{\rm in} = 50 {\rm AU}$ and $r_{\rm out} = 150 {\rm AU}$ to 
calculate the mass
of the ring part of the disk. The mass of the dust ring in model 1 (gas
mass $M_{\rm g} = 10 M_{\earth}$) is $9.0 \times 10^{-1} M_{\earth}$ and
the dust mass in model 2 ($M_{\rm g} = 1 M_{\earth}$) is $1.0 \times
10^{-1} M_{\earth}$. The gas to dust ratio is about 10. Thus, the gas-dust
coupling affects the motion of the gas only slightly, at best. (If the dust
grains settle down in the mid-plane of the disk, the dust density may
exceed the gas density. Even in this case, the dust grains do not affect
the gas motion strongly, because of the corotation of gas and dust in
steady orbits.) 

The above estimate of the mass assumes that the dust ring is composed of
grains smaller than $s \sim 200 \micron$.  Augereau et
al. (1999b) suggested the presence of meter-sized
bodies based on the spectral energy distribution of the HR 4796A dust
disk. The gas drag and radiation pressure force have little effect on
such large bodies, so unlike for the dust component,  the distribution
of these bodies is not strongly tied to the distribution of gas. The
presence of large meteorites and planetesimals (which do not fall under
the category of dust) necessarily increases the total mass of the 
disk of solid matter above our estimate of the dust disk mass. Additional
calculations are needed to model the long-term evolution of the large
bodies in disks. 

The total luminosity of thermal emission and scattered light from the
dust ring is
\begin{equation}
L_{\rm d} = \frac{\pi}{2} L \int_{r_{\rm in}}^{r_{\rm
out}}  \frac{s_{\rm s}^2 h_{\rm d} n_{\rm s}}{r} dr
= \frac{1}{2} L \int_{r_{\rm in}}^{r_{\rm out}} \frac{\tau}{r} dr \ .
\end{equation}
The above equation assumes the extinction cross section to be geometrical. The
luminosity of the dust ring in model 1 is $1.7 \times 10^{-2} L$
and that in model 2 is $2.0 \times 10^{-3} L$. 

The shape of the dust ring depends on the sharpness of the edge of the gas
disk (Fig. 9$c$). If the edge is more gradual, the ring is wider and its peak 
density is lower. However, the mass and luminosity of the ring do not strongly
depend on the sharpness of the disk edge. In model 5, where the width of
the edge is about 2 times wider than that in model 1, the mass and
luminosity of the ring are $M_{\rm d} = 7.2 \times 10^{-1} M_{\earth}$
and $L_{\rm d} = 1.4 \times 10^{-2} L$, respectively, which are only
about 20\% smaller than those in model 1. 

We can compare the above numbers with the masses of gas and dust
in transitional disks.
HR 4796A dust ring has been observed by thermal emission (Jayawardhana
1988; Koerner et al. 1998; Telesco et al.
2000) and by scattered light (Schneider et al.
1999). The emission from the dust ring has a peak
around 70 AU 
(Schneider et al. 1999; Telesco et al. 2000). 
The luminosities of thermal emission and scattered light are $L_{\rm th}
\simeq 
5 \times 10^{-3} L$ (Jura 1991) and $L_{\rm sc} \simeq 2
\times 10^{-3} L$ (Schneider et al. 1999), respectively. The
total luminosity of the ring is $L_{\rm d} \simeq 7 \times 10^{-3} L$. A
similar dust distribution with a peak at 72 AU and luminosity $L_{\rm d}
\simeq 
8 \times 10^{-3} L$ is found in our model 7, where the gas disk has a smooth
outer edge at 80 AU and mass $M_{\rm g} = 4 M_{\earth}$.  The gas mass
in model 
7,  $M_{\rm g} = 4 M_{\earth}$, is consistent with the observational upper
limit $M_{\rm g} < 7 M_{\earth}$ (Zuckerman et al. 1995;
Greaves et al. 2000).

The luminosities of the thermal emission and scattered light of HD
141569A are $L_{\rm th} \simeq 8 \times 10^{-3} L$ (Sylvester et
al. 1996) 
and $L_{\rm sc} \simeq (2-4) \times 10^{-3} L$ (Augereau et al.
1999a; Weinberger et al. 1999), 
respectively. 
The total luminosity of the dust is $L_{\rm d} \simeq 1 \times 10^{-2} L$.
The model which reproduces the dust ring with a peak
density at 325 AU and luminosity $L_{\rm d} \simeq 1 \times 10^{-2} L$ has
the gas disk with an edge at 370 AU and mass $M_{\rm g} = 50 M_{\earth}$
(model 8).
The inferred gas mass is consistent with the mass $M_{\rm g} \simeq
20-460 M_{\earth}$ derived by the observation (Zuckerman et
al. 1995). 

Although the gas masses derived by our models are consistent with the
observations, we must note that any exact match could be accidental. Our
models 
have large uncertainties. In particular, the efficiency of the collisional
destruction $K$ has an order of magnitude of uncertainty (eq. 
[\ref{eq:effdest}]). If $K$ of the actual grains were larger/smaller than the
value we adopted ($K=1.0 \times 10^{-9}$), then the grains would be more/less
easily destroyed and  the mass and luminosity of dust in a given gas disk
would be lower/higher. 

\subsection{Model Images of Scattered and Thermal Radiation}

In Figure 10, we present the theoretical image of model 4 (constant gas
surface density), inclined to 
the line of sight by 60$^\circ$ from pole-on, in the near-IR
scattered light (gray-scale brightness map at 1.1 $\micron$ wavelength),
and the mid-IR thermal radiation (over-plotted contours
of the 18.2 $\micron$ flux). Our modeling is somewhat simplistic, in that  the
scattering function is assumed to be isotropic (which does not alter the
general morphology, just removes any up-down asymmetry of the theoretical
image). However, the temperature profile as a function of radius  needs to be
computed as accurately as possible for a reliable prediction of thermal
IR flux distribution,  so we compute it within Mie theory, from  the
assumed properties of grains (specified earlier in \S2). 

The dust morphology in the model is qualitatively similar to those inferred
from observations  of the transitional system HD 141569A. The $18.2 \micron$
flux is centrally concentrated (like that imaged by Fisher et al. 2000), 
while the scattered light shows two separate components: the inner disk and
the outer ring, with the brightest parts being the  two projection-enhanced
ansae.  If these two components are  largely overlapping, they may create
the impression of  a shallow gap, resembling qualitatively the image  at
$\lambda \approx 1.1 \micron$ by Weinberger et al. (1999). [However, see 
Augereau et al. (1999a) for the dust distribution from their observations at
$\lambda \approx 1.6 \micron$.] The model disk in Figure 10 has a constant
gas surface density with a downturn beyond 100 AU. Model disks with surface
density decreasing with radius (model 1 etc.) show much more prominent inner
disks both in scattered light and in thermal emission.  Comparison with the
observations suggests that gas is not concentrated in the inner part of the
disk. The flat profile of gas surface density might arise after the viscous
evolution.

In order to achieve a quantitative agreement of our models with the
observations of HD 141569A, we would need to:  (i) consider non-isotropic
scattering function, (ii) allow for a more gradual outer gas truncation, and
(iii) convolve the theoretical images with the appropriate  point-spread
functions. As already mentioned, we defer the comprehensive modeling of
individual objects to a subsequent paper. 

Another system which might be qualitatively described by our  dust migration
models, HR 4796A, shows an enhanced-brightness ansae of the dust ring in
scattered light (Schneider et al. 1999), similar to our model in Figure 10.
The $18.2 \micron$ images also indicate a ring-like morphology (Jayawardhana
et al. 1998; Koerner et al. 1998; Telesco et al. 2000). In other words, the
dust ring with characteristic radius of 70 AU,  producing the scattered
light, also produces most of the $18.2 \micron$ thermal radiation. 

Our thermal mid-IR map shows a more centrally concentrated  morphology,
therefore our modeling procedure needs to be argued in  order to fit the
observations. Two possibilities exist.  First, our  model image represents
the emission from dust in steady circular orbits,  and omit the possible 
contribution by other particles, those flowing out on spiral orbits due to
$\beta > 1$, and the ones pumped  up to high eccentricity at the edge of the
disk (see \S5.2). The combined area of such dust grains is small, but
they could
in principle be seen because of the small physical sizes, and the associated 
higher temperatures (see Fig. 4 in Augereau et al. 1999b and Fig. 5 in Wyatt
et al. 1999). We note that the thermal emission in mid-IR (the Planck
function in Wien regime) and thus the visibility of grains in the outer ring
is often  very sensitive to the dust grain temperature; it is not uncommon 
to find that the predicted $18.2 \micron$ flux from the ring becomes 6 times
under just $\sim$20\% increase in dust temperature (from 70K to 84K), which
could make the  model ring visible.  Another way to assure a quantitative
fit  to HR 4796A data is a modification of the modeled gas profile, which
depletes the gas inside the ring radius. Our model would then produce a more
sharply defined ring and less dust inside the ring, because, at the region
where gas is depleted, collisional disruption of dust grains is violent. This
depletion of gas (and dust) might in this case be due to one or more planets
or a low-mass companion inside about a half of the ring radius ($\sim 40$AU).


\section{Discussion and Conclusions}

\subsection{ Dynamics of Dust Migration}

We have studied in the present paper the combined action of gas drag 
and radiation pressure on dust grains in optically thin circumstellar
gas disks.
Our results on dust migration are as follows.

1. Grains can migrate inward or outward, depending on whether its
azimuthal motion is faster or slower than the rotation velocity of gas.
Even grains smaller than the blow-out radius ($\beta > 1$) have
spiral trajectories in gas disks. 
   
2. We have provided analytical formulae for the radial migration speed
in \S3, which agree with the detailed numerical integrations.
The radial migration speed depends on gas density, but could be 
large. For grain sizes $s \lesssim 100 \micron$ in disks around A-type stars
of radius comparable with 100 AU and masses of order
$1-10$ Earth masses it reaches a significant fraction of gas sound speed. 

3. At some radius in the gas disk, a particle may reach a stable equilibrium
orbit where it corotates with the gas. Equilibrium is always found
outside the radius where the gas pressure is maximum. 

4. The radius of the steady orbit depends on the grain size. That
dependence is given by the rotation curve of  the gas disk, depending on
the radial gas distribution.
Smaller grains normally find a steady orbit farther from the
central star. This results in the fractionation of grains according to their
sizes.

5. The eccentricity of a grain is strongly damped through out most of the disk, 
but is excited and maintained at a value up to 
$e=(\beta - \eta)/(1-\beta)$ at the edge of the gas disk, where 
the gas density falls rapidly (e.g., as a steep power law or exponentially).
The rule number 3 above applies then, not to the whole orbit, but 
to the vicinity of the pericenter of the orbit. 

6. We have studied gas disk models which have density  varying as a power
law, truncated within an adjustable scale length,  always preserving the
Rayleigh stability of the disk, and a total  mass from a fraction to dozens
of Earth masses.  The outcome of grain migration depends on the radiation
pressure  coefficient $\beta$  and the grain size\footnote{ The primary
dependence is on $\beta$. The approximate ranges of grain radii are cited
below for $\beta(s)$ based on our physical model of grains as moderately
porous mixture of  olivines and pyroxenes, and for the luminosity to mass
ratio of the  star equal 8.4, as appropriate for HR 4796A and  approximately
also for HD 141569A.}. The fine dust fraction ($\beta \gtrsim 0.5$; $s
\lesssim 5 \micron$)  is removed by strong radiation pressure. Dust grains
with $0.1 \lesssim \beta \lesssim 0.5$,  or alternatively
$5 \lesssim s \lesssim 20 \micron$,
form an extended, outer dust disk component, residing most of the time beyond
the bulk of the gas disk on eccentric orbits.  Grains of intermediate size
($0.01 \lesssim \beta \lesssim 0.1$ or $20 \micron \lesssim s \lesssim 200
\micron$)   concentrate near the outer edge of the gas disk, forming a
prominent dust ring of width similar to the scale length of gas density
cutoff. Sand grains and meteoroids  ($\beta \lesssim 0.01$ or $s \gtrsim
200\micron$)  stay in the inner part of the gas disk. 

\subsection{Dust Morphologies and Their Observability}

We considered collisions between each grain size fraction.
The steady orbits can be called parking orbits at best.
After one collisional time (which is typically much
longer than the migration time, but much shorter than the age of the system)
the grains are fragmented or eroded, jumping into  a smaller grain
fraction and moving to a new parking orbit  or out into the interstellar
space. The grains are continuously replenished, 
in our  models from the inner part of the disk at radii of
order 10 AU, although a wider distribution of sources and/or somewhat 
different size distribution produced by dust source are certainly thinkable.
As a result, the disk reaches a stable dust density (and size) distribution. 
Including those particles which occupy the long-lived parking orbits, we
predicted the dust distribution in several model disks. The results show an
outer ring of small dust and, in addition, an inner disk
of large dust and sand.
These results are somewhat sensitive to the mechanical properties of the
particles, such as strength against catastrophic fragmentation and erosion
coefficient.  
In contrast, the dust density in Figure 9 does not depend on the 
dust production rate in the disk. 

A variety of dust morphologies can be obtained in our models with deliberately
smooth (monotonic) gas distributions, depending on how wide and dense the
outer ring is, relative to the main disk. For instance, there might exist
systems with ring morphology, ring plus inner disk morphology (possibly giving
impression of disks with gaps),  and disks with weak/unobservable rings.  Since
it is straitforward to obtain a ring several times denser than the adjacent
disk region, such rings should be observable in  scattered light imaging.
Somewhat more model-dependent is the observability of radial structure in
thermal IR maps, since  those depend strongly on the profile of dust
temperature as a function of radius. 

We have qualitatively 
compared our models with two observed transitional disks, around HR 4796A
and HD 141569A (by transitional we mean object intermediate between the
gas-rich primordial  nebulae and late-stage, replenished dust disks of
$\beta$ Pic and Vega-type;  cf.\ the discussion of this notion in the
introduction).
We found a qualitative agreement with most of the observed characteristics of
these disks, and sketched in \S7.2 several  possible improvements to the
modeling procedure, which could be used in detailed model fitting.

In general, our results open the prospect for diagnosing the state of the  gas
component of circumstellar disks by observing the dust and calculating the
dust-gas coupling.  The structure of the dust disk strongly depends on the
properties  of the gas disk. The most obvious example may be the existence,
radius and  width of the outer "edge" or downturn of the gas disk, reflected in
the corresponding  parameters of the dust ring (of course, what we call a
"ring" may well be in-out asymmetric, and thus prompt a description as a
truncated disk).  Therefore, if our model assumptions are satisfied, the dust
component of an observed disk carries valuable  information about the gas
component, which is often more difficult or impossible to study directly. For
example, unlike in HD 141569A, where the hydrogen and CO gas is at least
directly detected if not mapped in  detail, in the HR 4796A disk we only know
the upper limits on gas contents. 

\subsection{Origin of Structure in Disks: 
Dust Migration vs.\ Extrasolar Planets}

Our results also bear on the timely issue of planet detection in extrasolar
systems. Radial and azimuthal structure can be induced in circumstellar disks
by the gravitational perturbation (also known as sculpting) due to planets or
other substellar  companions in or in the vicinity of disks (cf.\  Dermott et
al. 1994, Liou \& Zook 1999,  Wyatt et al. 1999, and  Ozernoy et al. 2000, for
gas-free disks and, e.g.,  Lubow et al. 1999 for gas
disks). Detection of such structure is one of the main diagnostic tools in
many of the future imaging projects at various wavelengths, including the
ambitious space interferometry missions. However, caution should be exercised
while interpreting disk morphology. There are many possible reasons 
unrelated to dynamical perturbations which could, in principle, masquerade as
planets (Artymowicz 2000). 

First, optically thick disks with radius-dependent thickness might cast
shadows on parts of themselves. Next example would be a recent collision of
planetesimals in disk. Though the cloud of dust released in such a collision
would not be observable, it might become observable if exponentially 
amplified in the so-called dust avalanche (Artymowicz 1997, 2000).  This
requires the optical thickness in the disk mid-plane larger than that in the
$\beta$ Pic disk (transitional disks satisfy this requirement, as a rule).
While admittedly rare and improbable, such a collision might produce dense
patches or sectors in a disk, which are non-axisymmetric and  time-dependent
on orbital timescale,  features similar to those offered by dynamical
sculpting. Another example is the dust migration, potentially much more
common, but fortunately  easier to distinguish from planetary influences. 

Dust migration is capable of sculpting the disks, with an important
difference from dynamical sculpting  that it cannot itself produce
non-axisymmetry (nor,  presumably, any observable time variability). It can,
however, provide an alternative explanation of some observed features,
especially the radially structured dust disks, rings, and apparent gaps in
disks. By carefully analyzing the axisymmetry (or lack thereof) of the 
underlying dust distribution, taking properly into account the   asymmetry of
the image due to the non-isotropic scattering function of a collection of
dust grains, the observers will be able to decide whether  or not our
intrinsically axisymmetric models apply to any observed system. If
axisymmetric model reproduce the observations,   to within the uncertainties
of the latter, and the inferred gas amount does not clash with any
observational constraints,  then the dust migration hypothesis is preferred
over the planetary  hypothesis by the virtue of being more  natural
(conservative). This is  especially true in systems where gas has been
independently detected.

It appears that a convincing planet detection by way of dust imaging should
include not just a non-axisymmetric (and  time-dependent) structure in the
disk, but a very specific one, closely  following the predictions of
numerical simulations of either  gas-free or gas-rich disks. 

\acknowledgments
This work was supported by grants from NFR (Swedish Natural Sci.\
Res. Council). P.A. acknowledges a computer grant from Anna-Greta and Holger
Crafoord Foundation. T.T. was supported by the Research Fellowship of the
Japan Society for the Promotion of Science for Young Scientists.


\clearpage

\begin{deluxetable}{crrrrrrrr}
\tabletypesize{\footnotesize}
\tablecaption{Model Parameters. \label{tbl-1}}
\tablewidth{0pt}
\tablehead{
\colhead{Model} & \colhead{Star} & \colhead{$M_{\rm g}(M_{\earth})$} &
\colhead{$r_{\rm out}$(AU)} & \colhead{$p$} &
\colhead{$C_{\rm out}$} & \colhead{$K$} & \colhead{$M_{\rm d}(M_{\earth})$} & 
\colhead{$L_{\rm d}/L$}
}
\startdata
1 & HR4796A & 10 & 100 & $-2.25$ & 1.05 &
 $1 \times 10^{-9}$ & 0.90 & $1.7 \times 10^{-2}$ \\
2 & \nodata & 1 & \nodata & \nodata & \nodata & \nodata &
 0.10 & $2.0 \times 10^{-3}$ \\
3 & \nodata & \nodata & \nodata & $-2.75$ & \nodata & \nodata &
 0.37  & $8.2 \times 10^{-3}$ \\
4 & \nodata & \nodata & \nodata & $-1.25$ & \nodata & \nodata &
 3.10  & $4.2 \times 10^{-2}$ \\
5 & \nodata & \nodata & \nodata & \nodata & 2.0 & \nodata &
  0.72 & $1.4 \times 10^{-2}$ \\
6 & \nodata & \nodata & \nodata & \nodata & \nodata & $2 \times 10^{-8}$ &
  0.05  & $9.1 \times 10^{-4}$ \\
7 & \nodata & 4 & 80 & \nodata & \nodata & \nodata &
 0.31 & $8.2 \times 10^{-3}$ \\
8 & HD141569A &  50 & 370 & \nodata & \nodata & \nodata &
  4.23 & $1.1 \times 10^{-2}$ \\
\enddata


\tablecomments{Blanks mean that the same value as model 1 is used. See
\S6.1 for the definition of $K$ and \S7.1 for $M_{\rm d}$ and $L_{\rm d}$.
}

\end{deluxetable}


\clearpage

\begin{figure}
\plotone{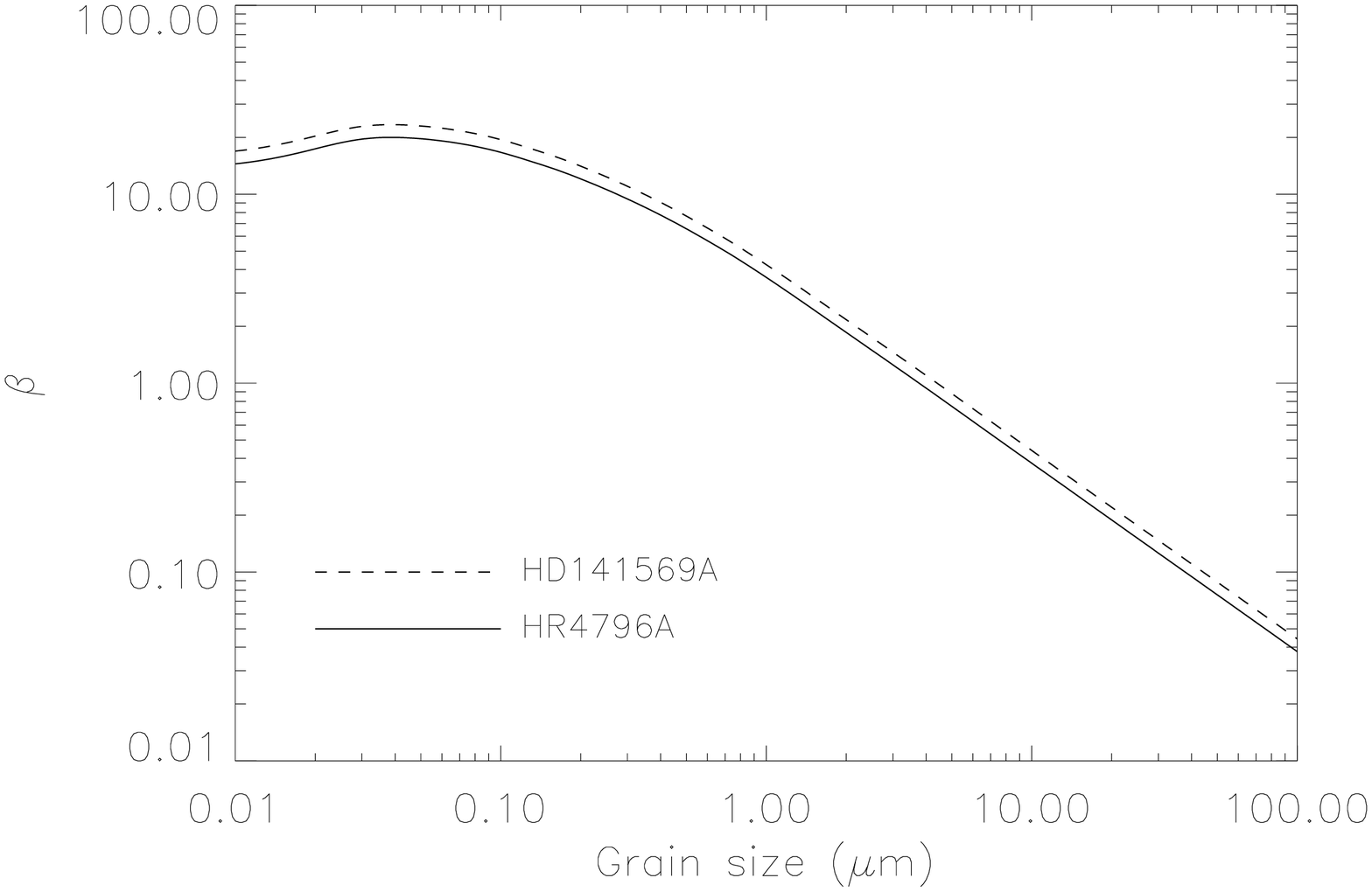}
\caption{
Radiation pressure to gravity ratio $\beta$ around HR 4796A (solid line)
and HD 141569A (dashed line).
\label{fig:beta}
}
\end{figure}
 
\begin{figure}
\plotone{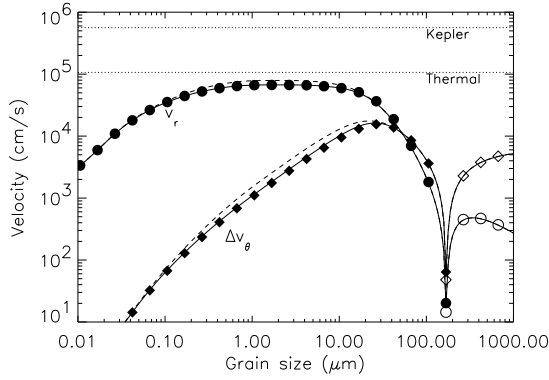}
\caption{
Velocities of dust grains at 70 AU in the gas disk of model 1 described
in \S4.
The circles show the radial velocities $v_r$
and the diamonds show the relative velocities in the azimuthal direction
between the grains and gas, $\Delta v_{\theta} = v_{\theta} - v_{\rm
g}$.
Filled symbols represent positive $v_r$ and negative $\Delta v_{\theta}$
(outward migration) and open symbols represent negative $v_r$ and
positive $\Delta v_{\theta}$ (inward migration).
The solid lines show the approximate formulae (\ref{eq:vt_connect}) and
(\ref{eq:vr_connect}) calculated numerically after the iterations for
the supersonic correction. 
The dashed lines show the first guesses of iteration, which are
calculated using the subsonic stopping time $T_{\rm ss}$ instead of
$T_{\rm s}$.
The horizontal dotted lines show the Kepler and thermal velocities at
70 AU in the model 1 disk.
\label{fig:velocity}
} 
\end{figure}

\begin{figure}
\plotone{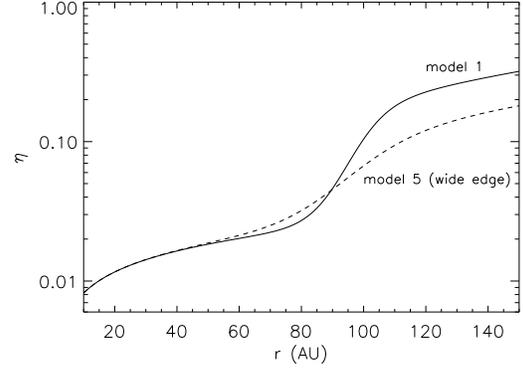}
\caption{
Pressure gradient force to gravity ratio $\eta$ in the disk of model 1
(soled line) and model 5 (dashed line).
The disk in model 5 has wider edge at 100 AU than that in model 1.
\label{fig:eta}
} 
\end{figure}

\begin{figure}
\plotone{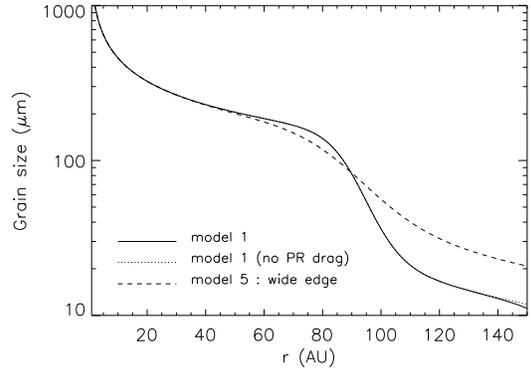}
\caption{
Size of grains on the steady orbits with radii $r$.
The solid and dashed lines show the grain size in model 1 and in model
5, respectively.
The dotted line represents the case where the Poynting-Robertson (PR)
drag is neglected in model 1, and shows the difference from the solid
line (including the PR drag) only for $r \gtrsim 130$AU.
\label{fig:size}
} 
\end{figure}

\begin{figure}
\plotone{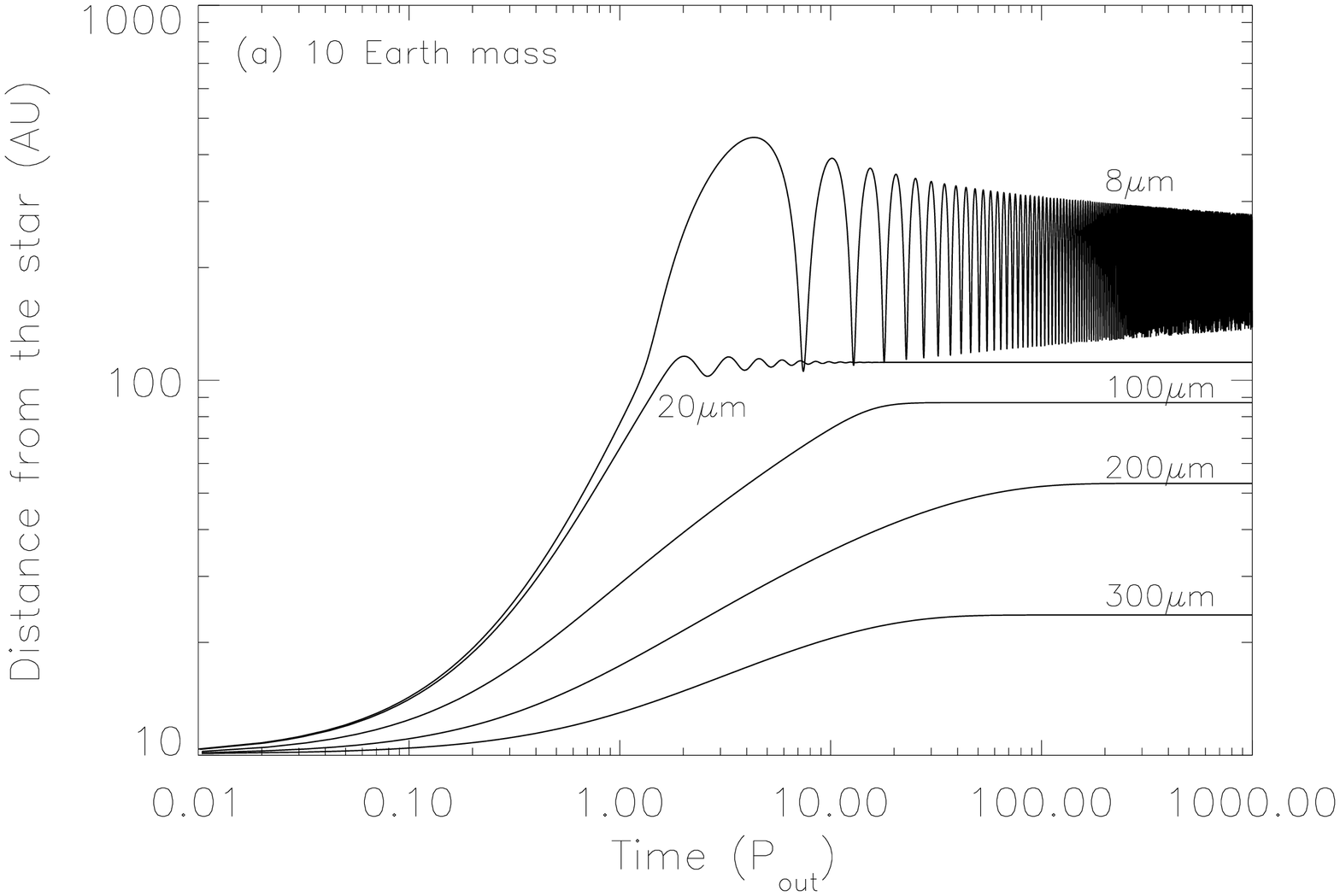}
\plotone{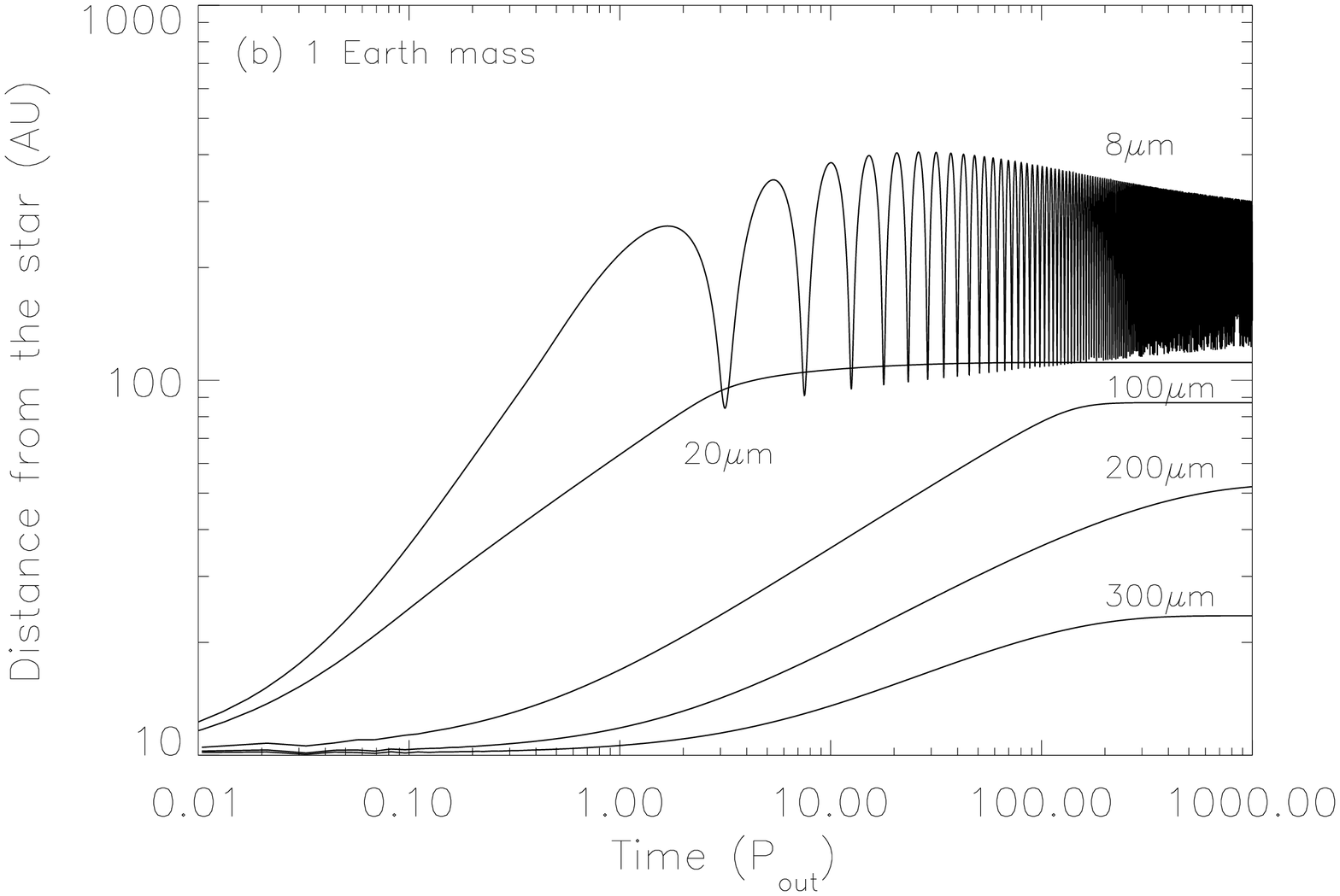}
\caption{
$(a)$ Evolution of the orbital radii $r$ of $8-300 \micron$ grains in
the disk of model 1.
The mass of the gas disk is $10 M_{\earth}$.
The time is measured by the orbital period at 100AU, $P_{\rm out}$.
The grains are generated at 10 AU and move outward until arriving at the
steady orbits.
The oscillations of $r$ of $8 \micron$ and $20 \micron$ grains are due
to the excitation of their orbital eccentricities at the edge of the gas
disk.
$(b)$ The same as Fig. 5$(a)$, but for model 2 where the gas disk is $1
M_{\earth}$.
\label{fig:radius}
} 
\end{figure}

\begin{figure}
\plotone{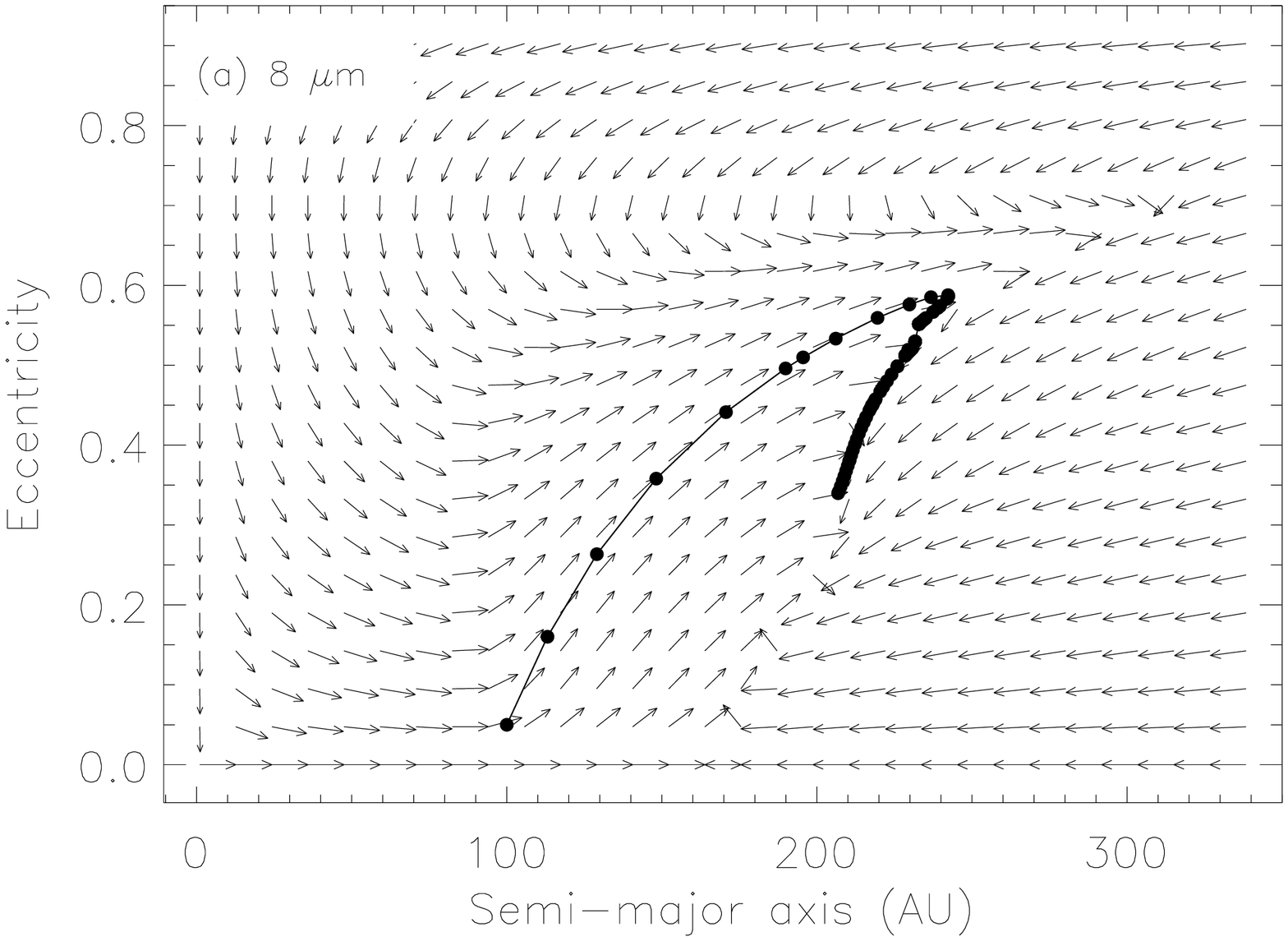}
\plotone{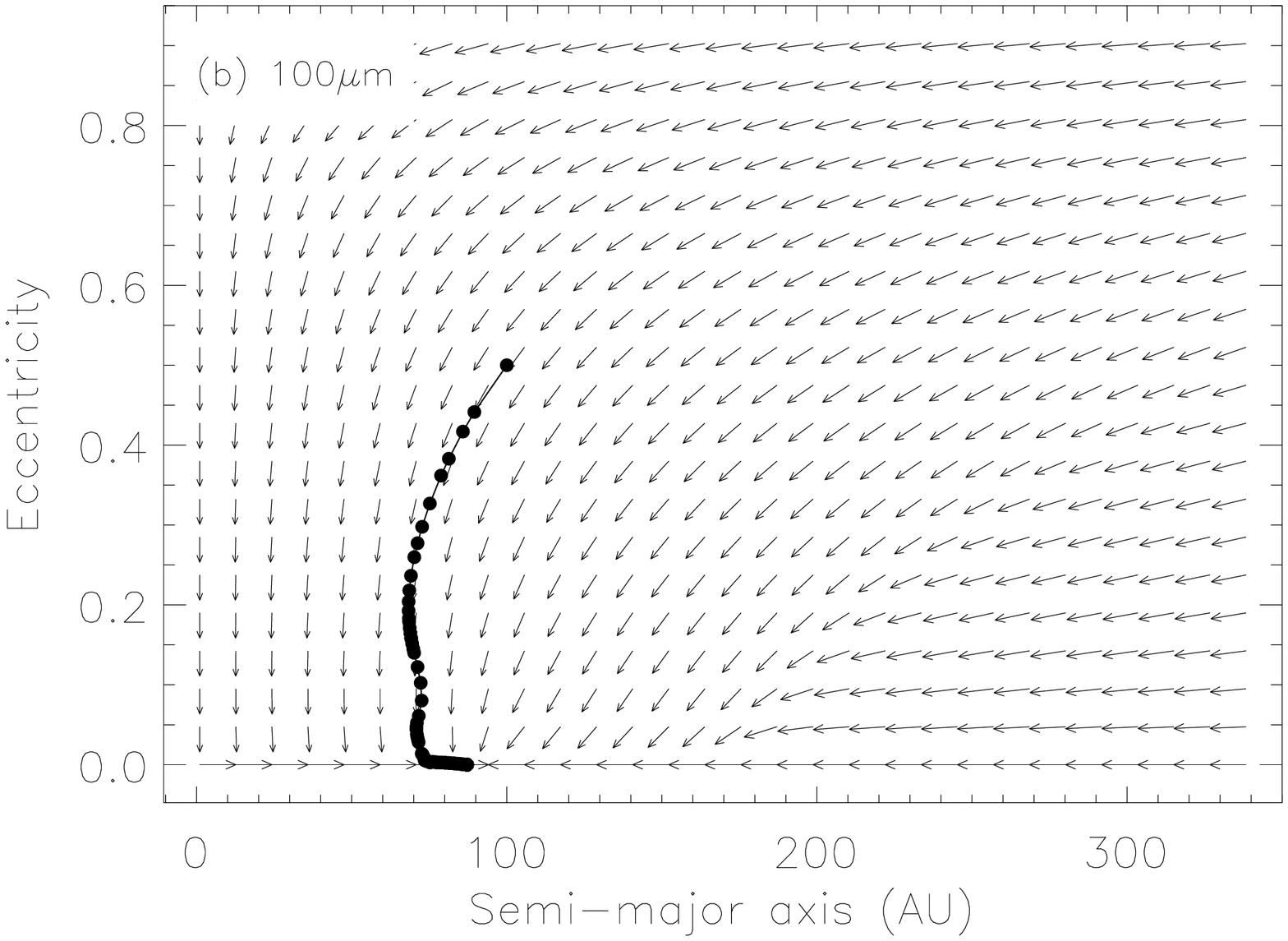}
\caption{
Evolution of the orbital elements of $(a)$ an $8 \micron$ grain and
$(b)$ a $100 \micron$ grain.
The gas disk is that in model 1.
Arrows show the directions of the orbital evolution obtained by the
perturbation equations (Gauss's equations). 
The length of arrows is set to be constant and does not represent the
evolution rate.
The circles show the evolution of orbital elements of test particles,
which are obtained directly by the orbital calculations.
The initial values of the test particles' orbital elements are $(a)$
$a=100$AU and $e=0.05$ and (b) $a=100$AU and $e=0.5$.
The circles are plotted until $t=1000 P_{\rm out}$ with the logarithmic
time interval: the first time interval is $0.01 P_{\rm out}$ and each
time interval is 1.2 times the previous one.
\label{fig:element}
} 
\end{figure}

\begin{figure}
\plotone{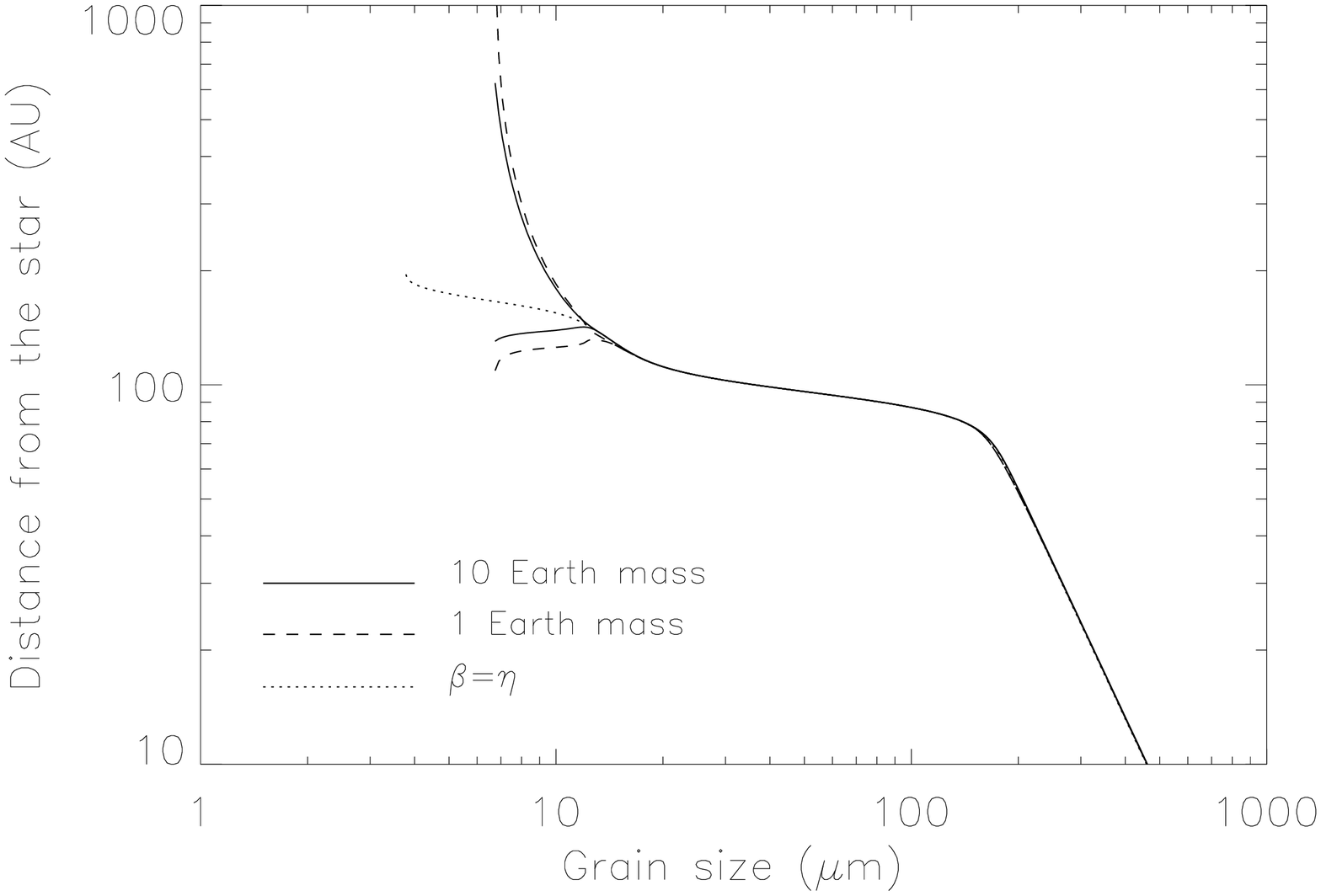}
\caption{
Distances of the apocenter and pericenter of the grains'
orbits from the central star.
The grains are generated at 10 AU and the orbits at $t=1000 P_{\rm out}$
are shown by the solid line (in the gas disk with $10 M_{\earth}$; model
1) and dashed line (in the gas disk with $1 M_{\earth}$; model 2).
The orbits of grains larger than about $10 \micron$ are circular and the
distances of the apocenter and pericenter coincide with the value
derived by the condition $\beta=\eta$ (dotted line).
\label{fig:apo-peri}
}
\end{figure}

\begin{figure}
\plotone{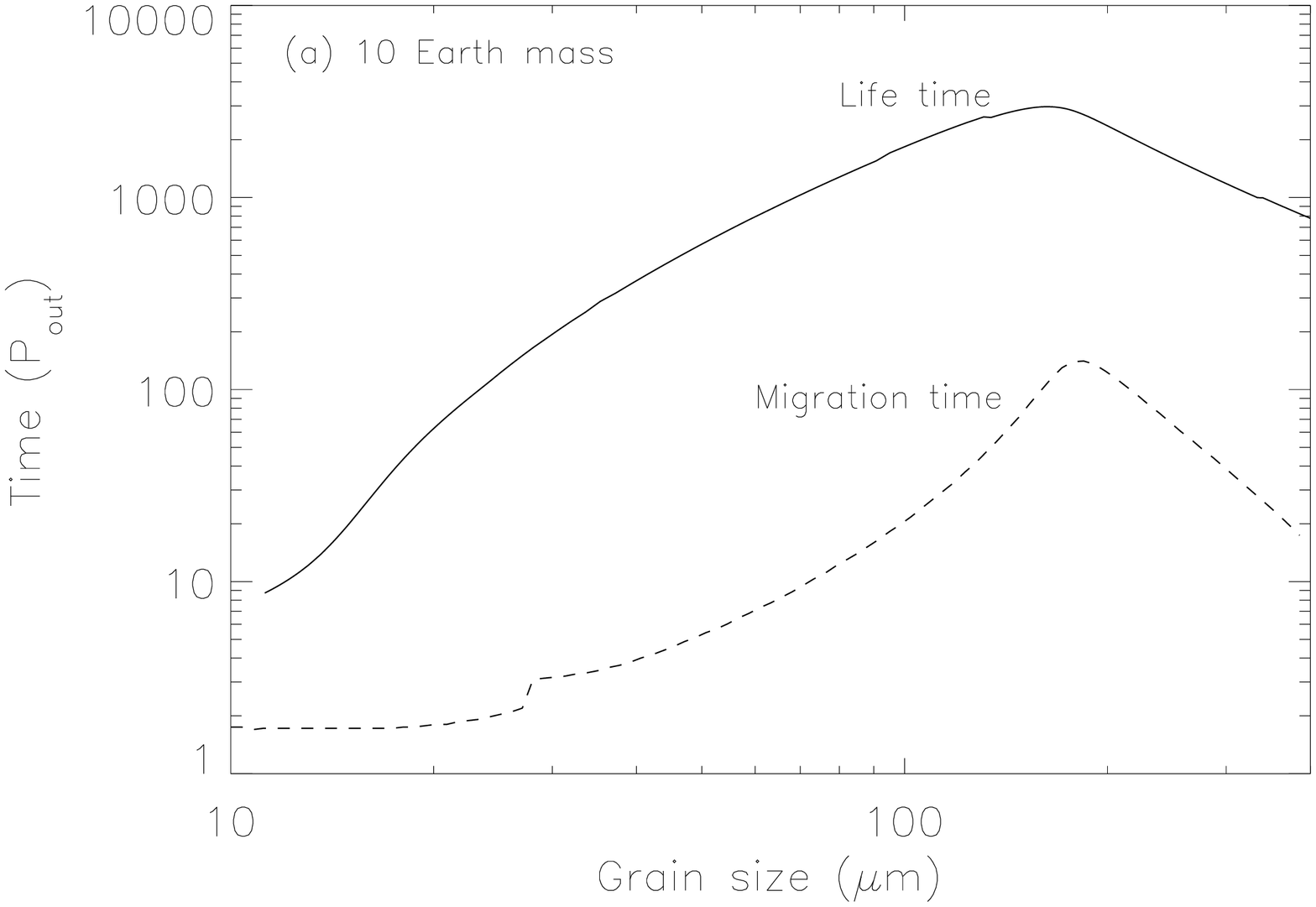}
\plotone{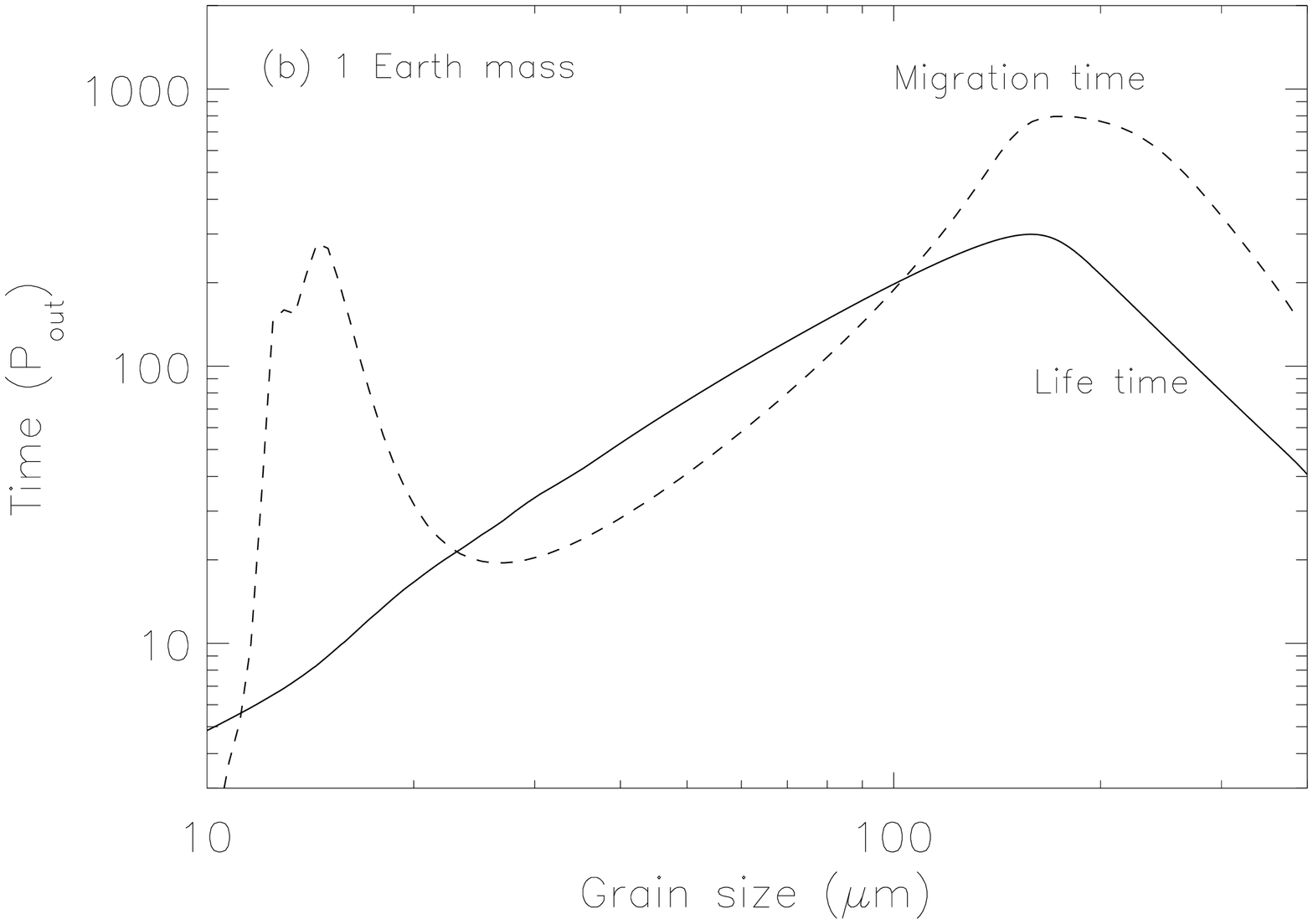}
\caption{
Life and migration time of dust grains. The solid line shows the life
time of grains on the steady orbits under the collisional destruction.
The dashed line shows the migration time from 10AU to the 99\% of the
radius of the steady orbits. $(a)$ The gas mass is $M_{\rm g} = 10
M_{\earth}$ (model 1). $(b)$ $M_{\rm g} = 1 M_{\earth}$ (model 2).
\label{fig:lifetime}
} 
\end{figure}

\begin{figure}
\plotone{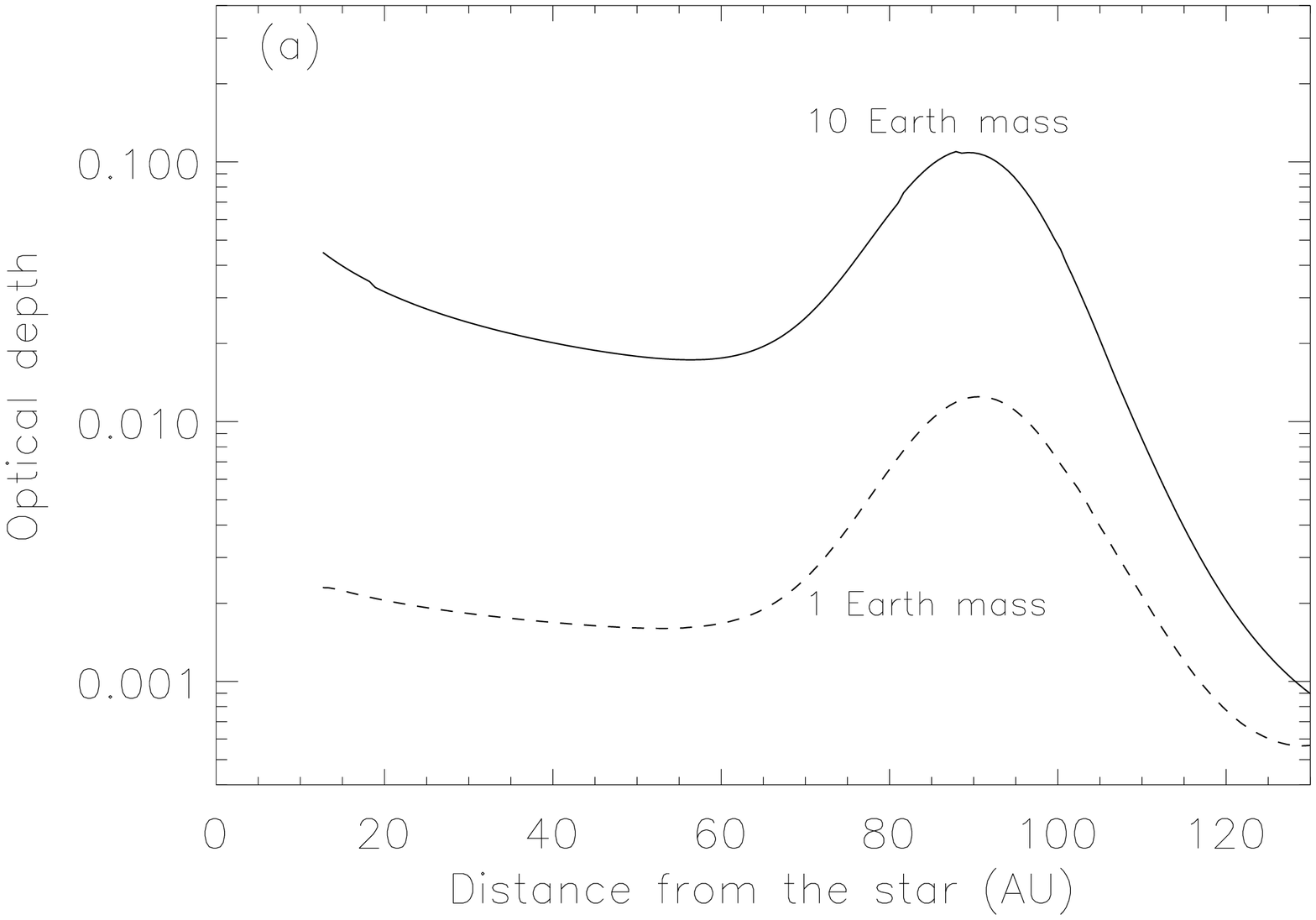}
\plotone{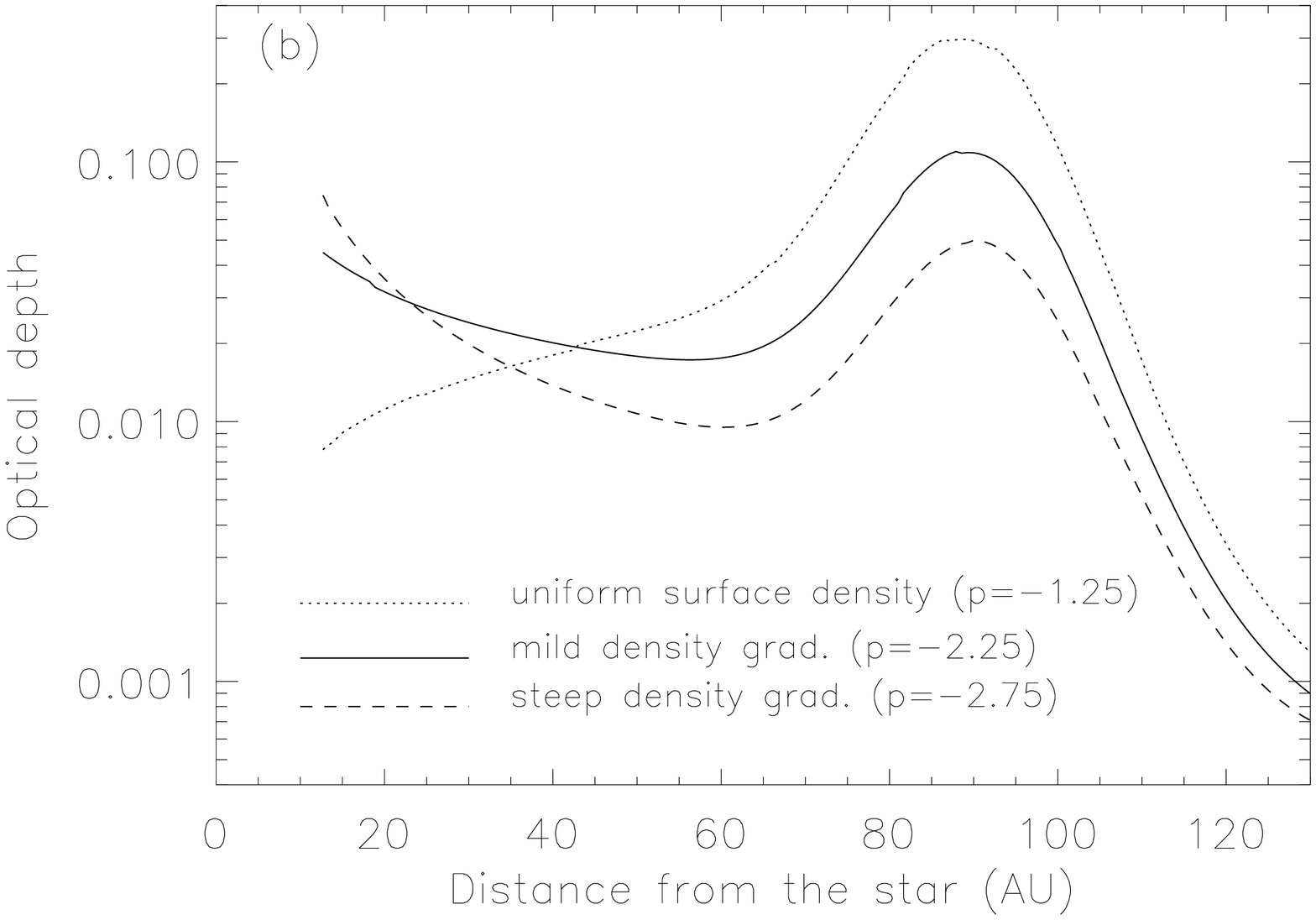}
\caption{
Optical depth in the vertical direction of the dust disks.
The optical depth is calculated using the geometrical cross sections of
grains.
$(a)$ For various disk masses, $M_{\rm g}$.
the solid and dashed lines show the optical depth of the dust in the gas
disk of $10 M_{\earth}$ (model 1) and $1 M_{\earth}$ (model 2),
respectively.
$(b)$ For various profiles of gas density, $p$, where $\rho_{\rm g}
\propto r^{p}$.
The dotted, solid and dashed lines correspond to $p=-1.25$ (uniform
surface gas density; model 4), $p=-2.25$ (model 1) and $p=-2.75$ (model
3), respectively.
$(c)$ For various width of the edge of gas disks.
The solid and dashed lines correspond to the narrow edge ($C_{\rm
out}=1.05$; model 1) and the wide edge ($C_{\rm out}=2.0$; model 5),
respectively.
$(d)$ For various strength of dust grains.
The solid and dashed lines correspond to the rocky grains ($K=1 \times
10^{-9}$; model 1) and the icy grains ($K=2 \times 10^{-8}$; model 6),
respectively.
\label{fig:tau}
} 
\end{figure}

\begin{figure}
\plotone{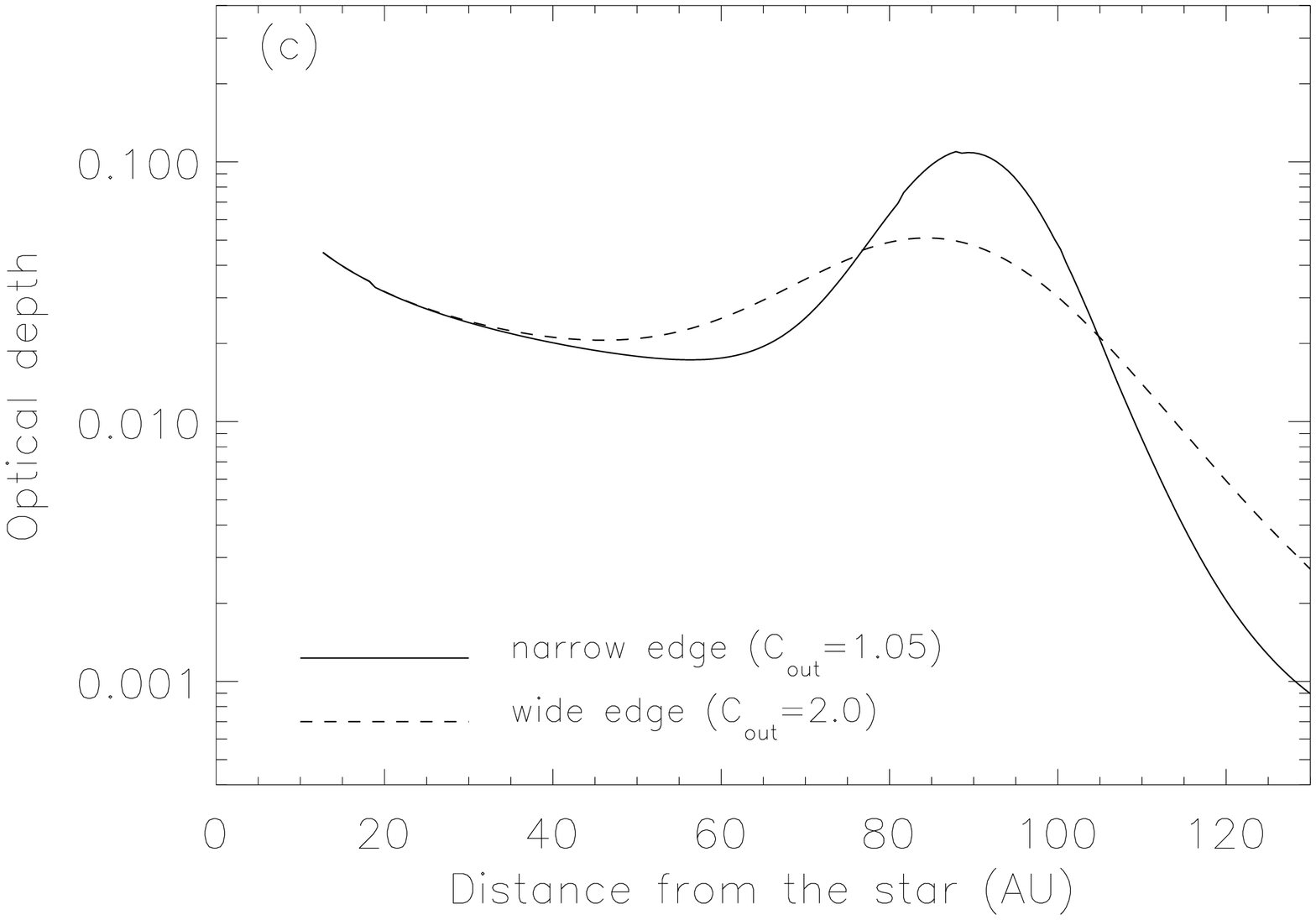}
\plotone{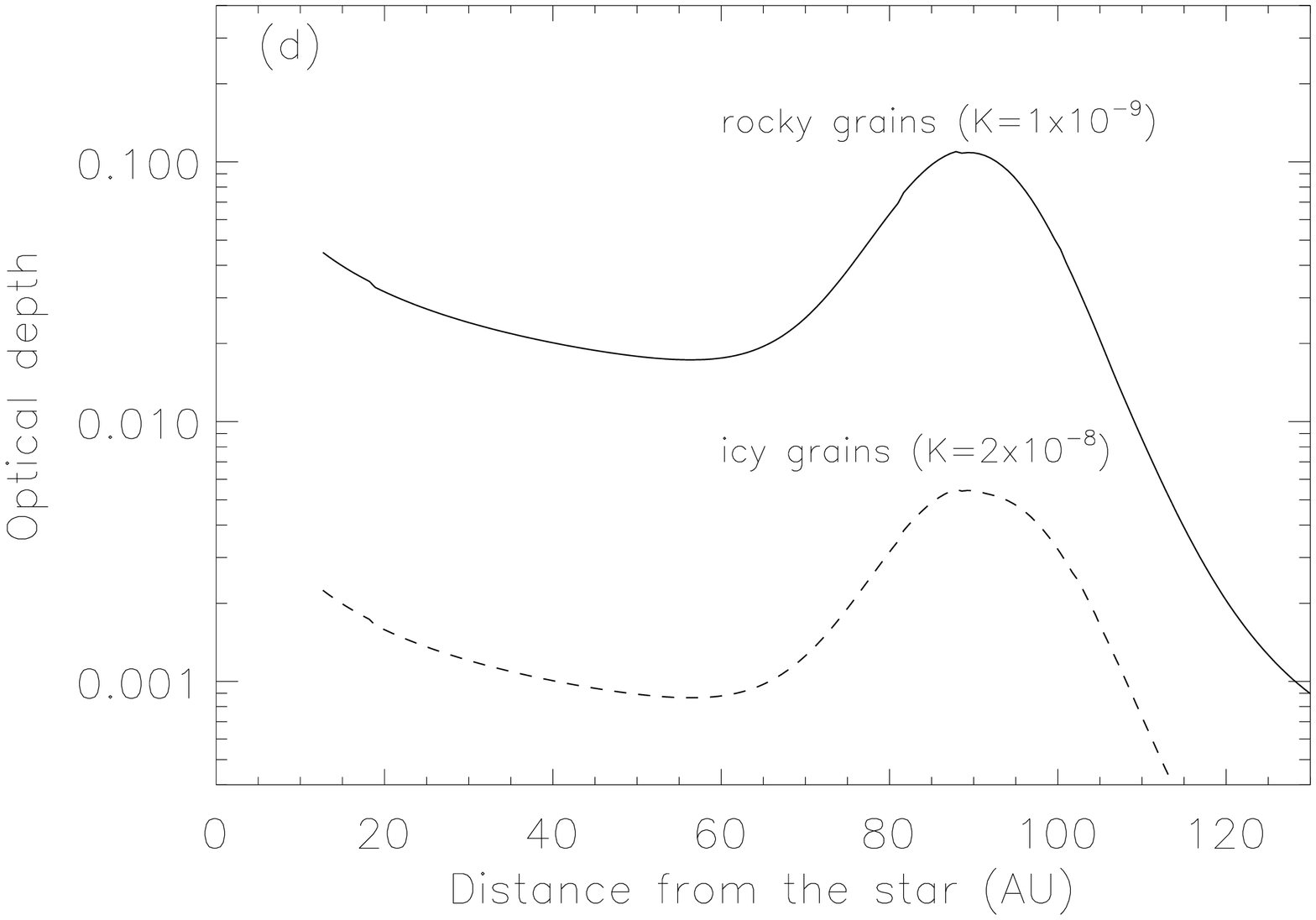} \\
Fig.9 --- continued.
\end{figure}

\begin{figure}
\plotone{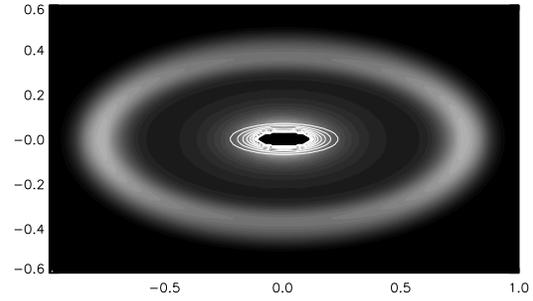}
\caption{
Simulated image of the model 4 disk in the scattered light and the thermal
emission. The gray scale shows the scattered light flux in $1.1 \micron$
with linear spacing. The contours show the thermal flux in $18.2
\micron$ and are spaced linearly. 
The unit of contour levels is arbitrary.
The disk has a constant surface gas density with an edge at 100 AU, viewed
in $60 \arcdeg$ from pole-on.
The flux is calculated for $r \ge 13$AU (outside the artificial hole at the
center). 
\label{fig:image}
} 
\end{figure}

\end{document}